\documentclass[fleqn,usenatbib]{mnras}

\usepackage{newtxtext,newtxmath}

\usepackage[T1]{fontenc}

\DeclareRobustCommand{\VAN}[3]{#2}
\let\VANthebibliography\thebibliography
\def\thebibliography{\DeclareRobustCommand{\VAN}[3]{##3}\VANthebibliography}

\usepackage{graphicx}	
\usepackage{amsmath}	

\renewcommand{\vec}{\mathbf}

\title[Cosmological multi-phase hydrodynamics]{Modeling multi-phase gases in cosmological simulations using compressible multi-fluid hydrodynamics}

\author[R. Weinberger and L. Hernquist]{
R. Weinberger$^{1}$\thanks{E-mail: rainer.weinberger@cfa.harvard.edu} and
L. Hernquist$^{2}$
\\
$^{1}$Canadian Institute for Theoretical Astrophysics, 60 St. George Street, Toronto, ON M5S 3H8, Canada\\
$^{2}$Center for Astrophysics | Harvard \& Smithsonian, 60 Garden Street, Cambridge, MA 02138, USA
}

\date{Accepted XXX. Received YYY; in original form ZZZ}

\pubyear{2021}

\begin{document}
\label{firstpage}
\pagerange{\pageref{firstpage}--\pageref{lastpage}}
\maketitle

\begin{abstract}
The diffuse medium in and around galaxies can exist in a multi-phase state: small, cold gas clouds contributing significantly to the total mass embedded in pressure equilibrium with a hotter, more diffuse volume-filling component. Modeling this multi-phase state in cosmological simulations poses a significant challenge due to the requirements to spatially resolve the clouds and consequently the interactions between the phases. In this paper, we present a novel method to model this gas state in cosmological hydrodynamical simulations. We solve the compressible two-fluid hydrodynamic equations using a moving-mesh finite-volume method and define mass, momentum and energy exchange terms between the phases as operator-split source terms. Using a stratified flow model, our implementation is able to maintain volume fraction discontinuities in pressure equilibrium to machine precision, allowing for the treatment of both resolved and unresolved multi-phase fluids. The solver remains second order accurate on smooth hydrodynamics problems. We use the source and sink terms of an existing two-phase model for the interstellar medium to demonstrate the value of this type of approach in simulations of galaxy formation, compare it to its effective equation of state implementation, and discuss its advantages in future large-scale simulations of galaxy formation.
\end{abstract}




\section{Introduction}
\label{sec:intro}

The theoretical understanding of the formation and evolution of galaxies has increased substantially over the last few decades.
Using constrains from the cosmic microwave background radiation and following the evolution of structure formation via gravitational interactions, we are able to predict the present-day large-scale structure distribution to a remarkable degree of accuracy \citep{schneider16}.
Including gas physics, radiative cooling and assuming a star formation efficiency for dense enough gas, it is possible to model the formation of the stellar component of galaxies from cosmological initial conditions \citep{springel03, schaye08, vogelsberger14}.
Recent studies, however, have shown that the process of gas cooling and star formation in galactic halos is more complex than a fraction of the gas being gravitationally \citep{jeans1902} or thermally \citep{field65} unstable. In particular, feedback effects from stars \citep{hopkins12b, gatto15, kim17} and active galactic nuclei \citep{wagner12, mukherjee18, cielo18}, as well as surface instabilities \citep{fielding20, gronke21} lead to substantial mass transfer between cold and hot gas phases, and a circulation in space as well as in thermodynamic state \citep{semenov17}.

In the interstellar medium, this leads to star formation processes down to AU scales impacting the future state of the galactic scale star-forming gas and even galactic winds on tens of kpc scales \citep{hopkins12}.
In the hot circumgalactic and intra-cluster medium, a similar problem of the co-existence of cold and hot phases of gas exists, with cold clouds fragmenting down to  sub-pc scales \citep{mccourt18}, while the characteristic scale of gaseous halos they are embedded in is several $10$s to $100$~kpc.
These enormous differences in scales poses a challenge to global simulations of galaxy formation. Even though increases in computational capability have made it possible to resolve galaxies \citep{dubois21, applebaum21, grand21} and the circumgalactic medium \citep{nelson16, hummels19, peeples19, vandevoort19, nelson20} in ever more detail even in cosmological simulations, the state of the art simulations are still by far not resolving relevant scales to accurately model the multi-phase gas in them.

Unable to cover all scales relevant to the problem, so-called sub-resolution models have to be employed to approximate the collective effect of the unresolved processes on resolved scales \citep[see][for a review]{somerville15}.
This, however, creates further problems: first, the averaged effect on larger scales is fairly uncertain, and second, the numerical implementation of these small scale effects into the employed hydrodynamics solver is non-trivial and a source of substantial modeling error by itself \citep{hopkins18, smith18, huang20, borrow21, huang22}.

In this paper, we present a novel way to overcome this problem by solving the equations of compressible two-fluid hydrodynamics instead of the regular single-fluid approach.
This allows us to split the underlying fluid equations into dynamics and source/sink terms and can directly include the complex spatially unresolved processes into the latter.
By choosing a multi-fluid model capable of precisely maintaining phase boundaries, we are in principle able to handle both, spatially resolved and unresolved multi-phase media or a mixture of both in the same simulation. For this first paper, however, we focus on applications in large scale simulations with unresolved phases.

In astrophysical simulations, hydrodynamics with multiple components are used in the case of a dust component moving relative to the embedded gas as in a protoplanetary disk \citep{benitez-llambay19}, or as a two-temperature electron-ion fluid in the case of AGN accretion disks \citep{sadowski17}.
In computational fluid dynamics outside astrophysics, the applications are more numerous \citep{prosperettiBook07}.
Here, we use one of these methods \citep{chang07} and modify it from its original application in liquid and gas mixtures to the above illustrated multi-phase gases of the interstellar, circumgalactic and intra-cluster media.

The paper is structured as follows: Section~\ref{sec:equations} introduces the equations of compressible multi-fluid hydrodynamics in an expanding spacetime and Section~\ref{sec:discretization} covers the discretization onto a moving, unstructured mesh using the finite-volume approach. The implementation is verified, and its performance illustrated with a number of test problems presented in Section~\ref{sec:verification}. Section~\ref{sec:application} covers its application to a number of commonly used galaxy formation simulations and their results are discussed in  Section~\ref{sec:discussion}. Finally, we summarize and conclude in Section~\ref{sec:conclusions}.

\section{Equations}
\label{sec:equations}

\subsection{General considerations}

Consider a mixture of 2 different fluid phases $j=\{1,2\}$ with density $\rho_j$, velocity $\vec{v}_j$, and specific thermal energy $u_{\text{th},j}$. Further, assume that the phases are distributed in sufficiently small droplets to justify the introduction of a volume fraction $\alpha_j$ filled by the respective phase\footnote{Note that this is a generalization, not a constraint, since $\alpha_i=1$ for a given phase $i$ yields the single-phase fluid relation. Importantly, however, it allows the different phases to have structure below an averaging scale.}. The mass of a given phase in a specific volume $V$ is thus
\begin{align}
    m_j &= \int\limits_{V} \alpha_j \rho_j dV.
\end{align}
Assuming each phase behaves in the same manner as a single-fluid except when exchanging mass, momentum or energy with the other phase, it is possible to derive volume-averaged 2-fluid equations (see e.g. \citealt{baer1986, saurel99}). This can be achieved by integrating the Euler equations over volume $V$ and separating the terms describing fluxes though the volume surface from internal fluxes between the phases \citep[][chapter 8]{prosperettiBook07}. The former combined with the time derivatives have the form of conservation laws of mass, momentum and total energy, while the latter are interaction terms that can be parameterized into source terms of mass exchange $\dot{m}$, drag forces $\vec{F}_d$ and heat transfer $Q$ between the fluids (assuming, for now, no external sources). These general 2-fluid equations are
\begin{align}
    &\partial_t \left(\alpha_1 \rho_1\right) +
    \nabla \cdot \left(\alpha_1 \rho_1 \vec{v}_1\right) = \dot{m} \label{eq:mfhydro1}\\
    &\partial_t \left(\alpha_1 \rho_1\vec{v}_1\right) +
    \nabla \cdot \left( \alpha_1 \rho_1 \vec{v}_1 \vec{v}_1^T + \alpha_1 p_1 \mathbf{I}\right) = \nonumber \\
    &\qquad p_i \mathbf{I} \nabla \alpha_1 + \dot{m} \vec{v}_i + \vec{F}_d \\
    &\partial_t \left(\alpha_1 E_1\right) + \nabla \cdot \left( \alpha_1 \vec{v}_1 \left(E_1 + p_1\right)\right) = \nonumber \\
    &\qquad p_i \vec{v}_i  \nabla \alpha_1 + \dot{m} E_i + \vec{F}_d \cdot \vec{v}_i + Q \label{eq:energy}\\
    &\partial_t \left(\alpha_2 \rho_2\right) +
    \nabla \cdot \left(\alpha_2 \rho_2 \vec{v}_2\right) = -\dot{m} \\
    &\partial_t \left(\alpha_2 \rho_2\vec{v}_2\right) +
    \nabla \cdot \left( \alpha_2 \rho_2 \vec{v}_2 \vec{v}_2^T + \alpha_2 p_2 \mathbf{I}\right) = \nonumber \\
    &\qquad p_i \mathbf{I} \nabla \alpha_2 - \dot{m} \vec{v}_i - \vec{F}_d \\
    &\partial_t \left(\alpha_2 E_2\right) + \nabla \cdot \left( \alpha_2 \vec{v}_2 \left(E_2 + p_2\right)\right) = \nonumber \\
    &\qquad p_i \vec{v}_i \nabla \alpha_2 - \dot{m} E_i - \vec{F}_d \cdot \vec{v}_i - Q \label{eq:mfhydro6}\\
    &\partial_t \alpha_1 + \vec{v}_i \nabla \cdot \alpha_1 = \dot{\alpha} \label{eq:alfaevol}\\
    &\alpha_1 + \alpha_2 = 1. \label{eq:alfasum}
\end{align}
The left side of equations~(\ref{eq:mfhydro1})-(\ref{eq:mfhydro6}) are identical to the Euler equations, only with $\alpha_j \rho_j$ as density, $\alpha_j \rho_j v_j$ as momentum density and $\alpha_j E_j = \alpha_j \rho_j u_{\text{th},j} + 1/2\,\alpha_j \rho_j \vec{v}_j^2$ as energy density (which they are in a volume-averaged sense). $p_j = (\gamma-1)\rho_j u_{\text{th},j}$ represents the pressure in the respective fluid. Additional non-conservative terms $p_i \mathbf{I} \nabla \alpha_j$ and $p_i \vec{v}_i  \nabla \alpha_j$ emerge in the momentum and energy equations, respectively, due to pressure forces caused by one fluid acting on the other fluid in the presence of gradients in volume fraction.
$v_i$, $p_i$, $E_i$ are interfacial velocity, pressure and energy density, which need to be modeled depending on the precise nature of the interaction between the fluids (e.g., if there is only mass flux from fluid 2 to fluid 1, none in the opposite direction and $\dot{m}>0$, the interfacial velocity $v_i = v_2$ to account for the appropriate momentum exchange between the fluids). Equations~(\ref{eq:alfaevol}) and (\ref{eq:alfasum}) describe the evolution of the volume fractions, with $\dot{\alpha}$ depending on the precise nature of the interaction between the fluids.

The key idea in this paper is to treat astrophysical multi-phase gases as 2-fluid systems with phase transitions being represented by $\dot{m}$, drag forces between clouds and their embedding medium by $\vec{F}_d$ and heat conduction between the phases by $Q$. The assumptions in the averaging procedure imply that the characteristic scale-length of clouds is much smaller than the averaging scale: the Field length \citep{begelman90}, cooling length \citep[e.g.][]{mccourt18} and/or mixing layers \citep{begelman90b, fielding20} need to be \textit{unresolved} for this formulation to be appropriate. In the opposite limit, where the cloud scale is fully resolved, each resolution element should contain only one phase. With these considerations in mind, we pay particular attention to the ability of the 2-fluid solver to be able to act as a quasi single-fluid solver in the case of fully resolved phase boundaries.

\subsection{Employed model in comoving coordinates}

In this work, we employ the above idea to the case of diffuse astrophysical gases in an extragalactic context. As such, it is critical to generalize the equations to comoving coordinates, as well as to include gravitational accelerations. In the following, we also make a number of simplifying assumptions: first, the equation of state for both fluids is identical, with $\gamma = 5/3$ for both fluids. While this is reasonable assumption for our application, it is substantially different from other applications of this type of 2-fluid modeling. This implies that we do not need to employ non-standard solvers due to Riemann problems with different equations of state at each side, simplifying the overall algorithm, but making comparisons with the literature difficult. Second, we assume the fluids to instantaneously reach pressure equilibrium,

\begin{align}
    p = \sum\limits_{j} \alpha_j (\gamma_j - 1) \rho_j u_{\text{th}, j} = (\gamma_j-1) \rho_j u_{\text{th}, j}.
\end{align}
which makes the volume fraction a function of the internal energy densities and eliminates the need to explicitly solve its evolution equation~(\ref{eq:alfaevol}). However, this assumption has a number of drawbacks: numerically, the equations become non-hyperbolic which is an issue for numerical stability of the scheme. In the literature \citep[e.g.][on which the presented implementation is largely based]{chang07}, an artificial interfacial pressure correction term is used to alleviate this problem. In the setups presented in this work we have not encountered stability problems, most likely due to the absence of a stiff equation of state and strong shocks, but we would expect the need for such a term in numerically more challenging setups. On top numerical considerations, the assumption of pressure equilibrium is not universally true in all situations we aim to model, and a gradual transition to pressure equilibrium (e.g., via the appropriate choice of $\dot{\alpha}$ in equation~(\ref{eq:alfaevol})) would be more appropriate. For simplicity, we nonetheless use the pressure equilibrium assumption in the following and leave improvements to future work.

Using the above-mentioned assumptions in the case of an ideal fluid in comoving coordinates with scale factor $a$ the 2-fluid equations are

\begin{align}
    &\partial_t \left(\alpha_j \rho_j\right) +
    \frac{1}{a}\, \nabla \cdot \left(\alpha_j \rho_j \vec{v}_j\right) = S_{\rho, j} \\
    &\partial_t \left(\alpha_j \rho_j\vec{\mathcal{V}}_j\right) +
    \nabla \cdot \left( \alpha_j \rho_j \vec{v}_j \vec{v}_j^T + \alpha_j p \mathbf{I}\right) = \nonumber \\
    &\qquad p \mathbf{I} \nabla \alpha_j - \frac{\alpha_j \rho_j}{a} \nabla \Phi + a \mathbf{S}_{v,j} \label{eq:momentum} \\
    &\partial_t \left(\alpha_j \mathcal{E}_j\right) + a \nabla \cdot \left( \alpha_j \vec{v}_j \left(E_j + p\right)\right) = \nonumber \\
    &\qquad - p \partial_t \alpha_j - \alpha_j \rho_j \left( \mathbf{v}_j \cdot \nabla \Phi \right) + a^2 S_{e, j} \label{eq:energy}\\
    &\sum\limits_j \alpha_j = 1
\end{align}
where we follow the notation similar to \citep{weinberger20}, i.e. $\vec{\mathcal{V}}_j = a \vec{v}_j$, $\mathcal{E}_j = a^2 E_j$ and $E_j$ denotes the total energy density of phase $j$. $\mathbf{I}$ is the unit-matrix. The terms $S_{\rho,j}$, $\mathbf{S}_{v,j}$ and $S_{e, j}$ denote sources and sinks due to mass exchange between the phases, drag forces and energy transfer (between phases) or radiative losses, respectively. Aside from these source terms, the most notable difference to the single-phase equations are the derivatives of $\alpha_j$ in the momentum and energy equations.

\subsection{Terms due to variable volume fractions}

The first term on the right-hand side in the momentum equation (\ref{eq:momentum}) can be intuitively understood by considering
\begin{align}
    \nabla\left(\alpha_j p \mathbf{I}\right) - p \mathbf{I} \nabla \alpha_j =  \alpha_j \nabla (p \mathbf{I}).
\end{align}
Thus, the additional term ensures no momentum flux due to pressure terms in the case of two fluids with constant pressure but spatially varying volume fractions. Maintaining this to machine precision is key to accurately modeling resolved phase boundaries and is non-trivial to achieve in the discretized version of these equations \citep[see discussion in][]{saurel99}.

The time-derivative term of the volume fraction in the energy equation (\ref{eq:energy}) represents the adiabatic energy change due to a change in volume fraction over time and can be taken care of during time-integration. Note, however that the volume fraction at a given point in time depends on the energy of the phases at that time. Keeping this consideration in mind, we now proceed to the discretization.

\section{Discretization}
\label{sec:discretization}

\begin{figure}
    \centering
    \includegraphics[width=\columnwidth]{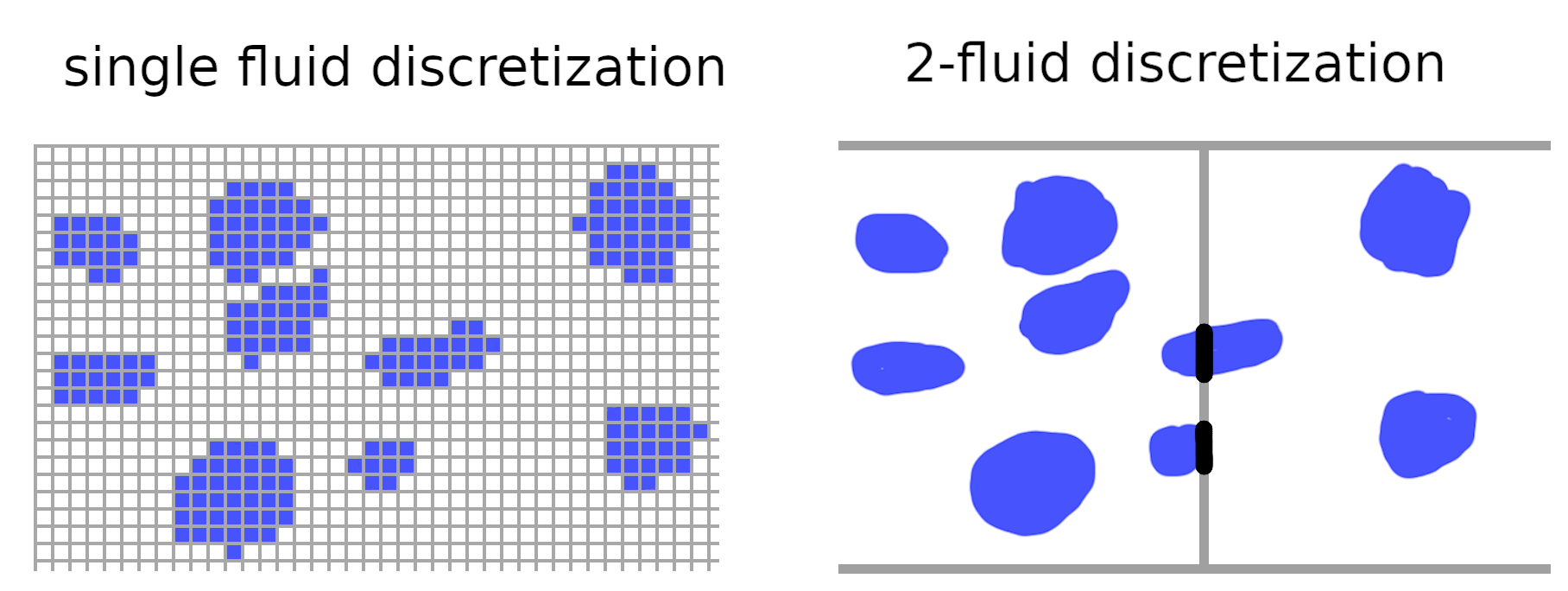}
    \caption{Schematic of a 2-fluid discretization as opposed to a single fluid discretization of a multiphase gas.}
    \label{fig:multifluid_schematic}
\end{figure}

We use the finite-volume, moving-mesh approach presented in \citet{springel10} to discretize the equations.
Changes in mass, momentum and total energy in a given cell are calculated by time-integrating the fluxes though all interfaces with neighbouring cells (ngb). In the following, it is useful to divide the flux into advective fluxes $\vec{F}_{kl}$ and pressure fluxes $\vec{G}_{kl}$ between cells $k$ and $l$. The advective flux density is given as
\begin{align}
    \vec{F}_{kl} &=
    \begin{pmatrix}
        \rho_{kl} \left(\vec{v}_{kl}-\vec{w}_{kl}\right)\cdot \hat{n}_{kl} \\
        \rho_{kl} \vec{v}_{kl} \left(\vec{v}_{kl}-\vec{w}_{kl}\right)^T\cdot \hat{n}_{kl}\\
        \rho_{kl} E \left(\vec{v}_{kl}-\vec{w}_{kl}\right)\cdot \hat{n}_{kl}\\
    \end{pmatrix}
    \label{eq:advflux}
\end{align}
where $\vec{w}_{kl}$ is the velocity of the interface between cells $k$ and $l$, and $\hat{n}_{kl}$ the normal vector of the interface. The pressure flux density is
\begin{align}
    \vec{G}_{kl} &=
    \begin{pmatrix}
        0 \\
        p \mathbf{I} \cdot \hat{n}_{kl}\\
        p \vec{v}_{kl} \cdot \hat{n}_{kl}\\
    \end{pmatrix}
    \label{eq:presflux}
\end{align}
The respective hydrodynamic properties at the interface are the solution to a Riemann problem between the two (spatially extrapolated) states of cells $k$ and $l$.
The time derivative of $\vec{Q} = (m, m\vec{v}, E\,V)^T$ in cell $k$ is then given by
\begin{align}
    \dot{\vec{Q}}_k = -\sum\limits_{l\in \text{ngb}} \left(\vec{F}_{kl}+\vec{G}_{kl}\right) A_{kl}
\end{align}
with $A_{kl}$ denoting the area of the interface between cells $k$ and $l$.

To generalize this to the multi-fluid equations, we use the stratified-flow model as presented in \citet{chang07}, and restrict ourselves to two phases in the following. The key idea behind the stratified flow model is that at each interface there are three different Riemann problems instead of a single one: One between phase $1$ on both sides, one between phase $2$ on both sides, and one between the two different phases. Figure~\ref{fig:multifluid_schematic} shows a schematic illustration of two 2-fluid cells and their interface illustrates these three cases.
The respective area fractions $\theta$ of each of these interfaces (again, between cells $k$ and $l$) are given by the overlap of the volume fractions on both sides
\begin{align}
    \theta_{1, kl} &= \min\left(\alpha_{k}, \alpha_{l}\right)\nonumber\\
    \theta_{2, kl} &= \min\left((1-\alpha_{k}), (1-\alpha_{l})\right)\nonumber\\
    \theta_{3a, kl} &= \max\left(0, \alpha_{k} - \alpha_{l}\right)\nonumber \\
    \theta_{3b, kl} &= \max\left(0, \alpha_{l} - \alpha_{k}\right) \label{eq:theta}.
\end{align}
Note that we use spatially constant volume fractions in each cell for simplicity. Here, $\alpha_k$ denotes the volume fraction of phase $1$ in cell $k$, thus $(1-\alpha_k)$ is the volume fraction of phase $2$, allowing us to drop the index $j$ connected to the phase. The last two lines in equation (\ref{eq:theta}) represent cases where in the cross-phase boundary, phase 1 is present in $k$ and phase 2 in cell $l$ (denoted with index $a$) or phase 2 in $k$ and phase $1$ in $l$ (index $b$).
The time derivative of  $\vec{Q}_1 = (m_1, m_1\vec{v}_1, \alpha_1\, E_1\,V)^T$ is then given by
\begin{align}
    \dot{Q}_{1, k} =& -\sum\limits_{l\in \text{ngb}} \theta_{1, kl} \left(\vec{F}_{1, kl} +\vec{G}_{1, kl} \right) {A}_{kl} \nonumber\\
    &- \sum\limits_{l\in \text{ngb}} \theta_{3a, kl} \left(\vec{F}_{3a, kl} + \vec{G}_{3a, kl} \right) {A}_{kl}  \nonumber\\
    &- \sum\limits_{l\in \text{ngb}} \theta_{3b, kl} \left(\vec{F}_{3b, kl} + \vec{G}_{3b, kl}\right) {A}_{kl}.
\end{align}

For the phase 1 - phase 1 interface, the respective flux density vectors $\vec{F}_{1, kl}$ and $\vec{G}_{1, kl}$ are defined as in the single-fluid case (using the solution to the first Riemann problem, $\rho_{1, kl}, \vec{v}_{1, kl}, p_{1, kl}$). The fluxes across phases use the solution to the third Riemann problem, $\rho_{3, kl}, \vec{v}_{3, kl}, p_{3, kl}$ and differ since in case $a$ there can only be advective outflow and in case $b$ only advective inflow:
\begin{align}
    \vec{F}_{3a, kl} &=
    \begin{pmatrix}
        \rho_{3, kl}\,\max(0,\left(\vec{v}_{3, kl}-\vec{w}_{kl}\right)\cdot \hat{n}_{kl})\\
        \rho_{3, kl} \,\vec{v}_{3, kl}\,\max(0,\left(\vec{v}_{3, kl}-\vec{w}_{kl}\right)\cdot \hat{n}_{kl})\\
        \rho_{3, kl} \,E_{3, kl}\, \max(0,\left(\vec{v}_{3, kl}-\vec{w}_{kl}\right)\cdot \hat{n}_{kl})
    \end{pmatrix} \\
    \vec{F}_{3b, kl} &=
    \begin{pmatrix}
        \rho_{3, kl}\,\min(0,\left(\vec{v}_{3, kl}-\vec{w}_{kl}\right)\cdot \hat{n}_{kl})\\
        \rho_{3, kl} \,\vec{v}_{3, kl}\,\min(0,\left(\vec{v}_{3, kl}-\vec{w}_{kl}\right)\cdot \hat{n}_{kl})\\
        \rho_{3, kl} \, E_{3, kl} \, \min(0,\left(\vec{v}_{3, kl}-\vec{w}_{kl}\right)\cdot \hat{n}_{kl})
    \end{pmatrix}.
\end{align}

The pressure fluxes are different for cases $a$ and $b$:

\begin{align}
        \vec{G}_{3a, kl} &=
    \begin{pmatrix}
        0\\
        p_{3, kl} \mathbf{I} \cdot \hat{n}_{kl}\\
        p_{3, kl} \max\left(0, (\vec{v}_{3, kl} -\vec{w}_{kl}) \cdot \hat{n}_{kl}\right) + p_{3, kl} \vec{w}_{kl} \cdot \hat{n}_{kl}
    \end{pmatrix} \\
        \vec{G}_{3b, kl} &=
    \begin{pmatrix}
        0\\
        0\\
        p_{3, kl} \min\left(0, (\vec{v}_{3, kl} -\vec{w}_{kl}) \cdot \hat{n}_{kl}\right)
    \end{pmatrix}.
\end{align}
Note that the momentum term of the pressure flux acts only on phase 1 in case a (it acts on phase 2 in case b). The energy term is analogous to the advection flux terms, however, with an additional one-sided term. As discussed in \cite{springel10} it is possible to use the lab-frame velocity $\vec{v}_{kl}$ for this component for all interfaces irrespective of their velocity. This simply leads to a constant nonzero flux due to this Galilean transformation. In our multifluid case, however, care has to be taken that this constant flux is not double-counted, thus the one-sided nature of this term. For a constant mesh $w_{kl}=0$, both terms are identical.

For $\vec{Q}_2 = (m_2, m_2\vec{v}_2, \alpha_2\, E_2\,V)^T$, we obtain

\begin{align}
    \dot{Q}_{2, k} =& -\sum\limits_{l\in \text{ngb}} \theta_{2, kl} \left(\vec{F}_{2, kl} +\vec{G}_{2, kl} \right) {A}_{kl} \nonumber\\
    &- \sum\limits_{l\in \text{ngb}} \theta_{3, a, kl} \left(\vec{F}_{3, b, kl} + \vec{G}_{3, b, kl} \right) {A}_{kl}  \nonumber\\
    &- \sum\limits_{l\in \text{ngb}} \theta_{3, b, kl} \left(\vec{F}_{3, a, kl} + \vec{G}_{3, a, kl}\right) {A}_{kl};
\end{align}
i.e. the role of cases a and b is interchanged in the flux terms.

Time integration as well as gradient calculation is done the same way as the single-fluid case \citep{pakmor16}.
The time derivative term in equation (\ref{eq:energy}) is subsequently discretized as
\begin{align}
    - \int p\partial_t \alpha dt \rightarrow p_n \alpha_n - p_n \alpha_{n+1} \label{eq:dtalpha} \\
    \alpha_{n+1} = \frac{H_\text{1, n+1}}{H_\text{1, n+1}+H_\text{2, n+1}}
\end{align}
where
\begin{align}
H_\text{1,n+1} &= (\rho_1 u_{th, 1} \alpha V)_n + (p \alpha V)_n + \int\limits_{n}^{n+1}{\dot{Q}_{1,3}} \,dt \\
H_\text{2,n+1} &= (\rho_2 u_{th, 2} (1-\alpha) V)_n + (p (1-\alpha) V)_n + \int\limits_{n}^{n+1}{\dot{Q}_{2,3}} \,dt
\end{align}
is the enthalpy at step $n+1$, calculated using the internal energy densities $\rho_1 u_{th, 1},\, \rho_2 u_{th, 2}$, volume fraction $\alpha$ and pressure $p$ of timestep $n$ and the numerically evaluated integrated energy fluxes ($\dot{Q}_{1,3}, \, \dot{Q}_{2,3}$) though cell boundaries from timestep $n$ to timestep $n+1$. Note that we make use of the fact that the term~(\ref{eq:dtalpha}) leaves the enthalpy constant which allows us to calculate the enthalpy at timestep $n+1$ without simultaneously needing to solve for $\alpha_{n+1}$. Using $H_\text{1, n+1}$ and $H_\text{2, n+1}$ we can calculate the volume fraction $\alpha_{n+1}$, and subsequently evaluate term~(\ref{eq:dtalpha}) to obtain the internal energy densities at timestep $n+1$. Note that we assume the same equation of state for both phases (as does the Riemann solver calculating the solution to interface 3), which is the main difference to \citet{chang07}.

The timestep constraint for the fluxes are given by
\begin{align}
    \Delta t = c_\text{CFL} \, r_\text{cell} / \max(c_{s,1}, c_{s,2})
\end{align}
for moving-mesh simulations and
\begin{align}
    \Delta t = c_\text{CFL} \, r_\text{cell} / \max(c_{s,1}+\left|\vec{v}_{1}\right|, c_{s,2}+\left|\vec{v}_{2}\right|)
\end{align}
for a fixed mesh. The CFL factor $c_\text{CFL}=0.3$ is kept fixed throughout this study, $r_\text{cell}$ is the radius a sphere with identical volume as the cell would have, $c_{s,1}$ and $c_{s,2}$ are the sound speed of fluid 1 and fluid 2, respectively. For a discussion of the mathematical properties of the equations, see \citet{saurel99}, section~3.3.

\subsection{Source terms}

The source terms are implemented in a Strang split fashion, with half-steps in between gravity and hydrodynamics terms \citep[see][for gravity and hydrodynamics operations, respectively]{springel10, pakmor16} using explicit time integration. For numerical reasons we time-integrate  $S_{u,1}$ the source term of the specific internal energy $u_{\text{th},1}$ instead of directly integrating $S_{e,1}$.
To ensure accuracy of the source-term integration, we require an additional timestep criterion on top of the traditional CFL criterion, and set the timestep (for both interface fluxes and source terms) to the more restrictive one:

\begin{align}
    \Delta t = 0.1 \, \min\left(\left|\alpha \rho_1S_{\rho, 1}^{-1}\right|, \left|(1-\alpha) \rho_2 S_{\rho,2}^{-1}\right|, \left|u_1  S_{u, 1}^{-1}\right|\right).
\end{align}

\subsubsection{Multifluid interstellar medium model}
In most of the present work we use the model presented in \citet{springel03} by directly using the differential equations that are their starting point (note the different notation).
\begin{align}
  S_{\rho, 1} &= \frac{(1-\alpha) \rho_2}{t_*} (\beta + A \beta) -\frac{1-f}{u_{\text{th},1}-u_{\text{th},2}} \Lambda(\rho_1, u_{\text{th},1}) \label{eq:src_rho1}\\
  S_{\rho, 2} &= -\frac{(1-\alpha) \rho_2}{t_*} (1+ A \beta) +\frac{1-f}{u_{\text{th},1}-u_{\text{th},2}} \Lambda(\rho_1, u_{\text{th},1}) \label{eq:src_rho2}\\
  S_{v, 1} &= (\beta + A \beta) \frac{(1-\alpha) \rho_2}{t_*} \vec{v}_2\\
  S_{v, 2} &= \frac{1-f}{u_{\text{th},1}-u_{\text{th},2}} \Lambda(\rho_1, u_{\text{th},1}) \vec{v}_1\\
  S_{e, 1} &= \frac{(1-\alpha) \rho_2}{t_*}\left[ \beta(u_\text{SN} + u_{\text{th},2}) + A \beta u_{\text{th},2}\right] \nonumber\\
  &-\frac{u_{\text{th},1} - f u_{\text{th},2}}{u_{\text{th},1}-u_{\text{th},2}} \Lambda(\rho_1, u_{\text{th},1}) \label{eq:src_egy}
\end{align}

Phase~1 is the hot phase and initially dominating ($\alpha=1-10^{-8}$ in initial conditions). Phase~2 is the cold phase and quasi-isothermal by setting the temperature to $1000$~K after each timestep.
The model has two different regimes controlled by variable $f\in\left[0,1\right]$, depending on the total density $\rho = \alpha \rho_1 + (1-\alpha)\rho_2$.
\begin{align}
    f = 0.5 + 0.5 \tanh\left( 10 (\log(\rho)-\log(\rho_\text{th}))\right),
\end{align}
with $f=1$ representing a single-phase mode on densities below a threshold, $\rho_\text{th}=1.039\times10^{-25}$~g~cm$^{-3}$, and $f=0$ a star forming, 2-phase mode.
Cooling of the hot phase is used in the single-phase mode, while the 2-phase mode assumes a transition from hot to cold phase described by the last term in eq. (\ref{eq:src_rho1}) and eq. (\ref{eq:src_rho2}).
The star formation rate density is defined as
\begin{align}
    \dot{\rho}_{*} = \frac{(1-\alpha) \rho_2}{t_*}
\end{align}
with the star formation timescale $t_*$ given by
\begin{align}
    t_* =
    \begin{cases}
        t_{*,0} \left(\frac{\rho}{\rho_\text{th}}\right)^{-\frac{1}{2}}\quad &\text{if } \rho > \rho_\text{th}, \\
        \infty \quad&\text{else.}\\
    \end{cases}
\end{align}
$t_{*,0}= 2.1$~Gyr is a free parameter of the model.
The model assumes that a fraction $\beta=0.1$ of the stars formed explode as supernovae instantaneously, their gas being added to the hot phase with a supernova temperature $T_\text{sn} = 5.73\times 10^7$~K (and the respective specific internal energy $u_\text{SN}$). The supernovae furthermore are assumed to evaporate a fraction cold gas, parameterized by the evaporation factor
\begin{align}
A = A_0 \left(\frac{\rho}{\rho_\text{th}}\right)^{-\frac{4}{5}},
\end{align}
which specifies the mass of evaporated gas per unit mass of exploded supernovae. We choose $A_0=573$.

The source terms in the momentum equations represent the momentum transfer due to mass exchange of the phases. Note that we do not include any drag terms or evaporation due to shear flows in this study.

As mentioned above, we integrate the rate of change of the specific internal energy rather than the energy density, which can be derived from eq.~(\ref{eq:src_rho1}) and eq.~(\ref{eq:src_egy})
\begin{align}
    S_{e,1} &= S_{\rho,1} u_{\text{th},1} + \rho_1 S_{u,1} \\
    S_{u,1} &= \frac{(1-\alpha) \rho_2}{t_*} \left[ \beta (u_\text{SN} + u_{\text{th},2} - u_{\text{th},1}) + A \beta (u_{\text{th},2} - u_{\text{th},1}) \right] \nonumber\\
    &+ f\Lambda(\rho_1, u_{\text{th},1}) \label{eq:src_u}
\end{align}

\subsubsection{Effective equation of state model}

We compare this on-the-fly solution of the source terms to the well-established effective equation of state model with the same underlying parameters. Within the effective equation of state framework, the single-fluid Euler equations are solved with ordinary cooling below $\rho_\text{th}$ and a density dependent pressure at higher densities. Defining $x=(1-\alpha) \rho_2 \rho^{-1}$, the effective equation of state pressure is given by
\begin{align}
    P_\text{eeos} = (\gamma - 1) \rho \left[ (1-x)u_{\text{th},1} + x u_{\text{th},2}\right]
\end{align}
 $u_{\text{th},1}$ is derived from an equilibrium assumption in the specific internal energy equation i.e. eq. (\ref{eq:src_u}) with $f=0$ vanishes, which leads to
 \begin{align}
     u_{\text{th},1} = \frac{u_\text{SN}}{A+1} + u_{\text{th},2},
 \end{align}
 while $u_{\text{th},2}$ is the specific internal energy corresponding to a temperature of $1000$~K. The mass fraction in cold clouds can be obtained by assuming that the pressure, or alternatively the combined internal energy density is constant in time, i.e. energy injection due to supernovae balances losses due to radiative cooling and star formation
\begin{align}
    \beta \frac{(1-\alpha) \rho_2}{t_*} u_\text{SN} = \Lambda(\rho_1, u_{\text{th},1}) + (1-\beta) \frac{(1-\alpha) \rho_2}{t_*} u_{\text{th},2}
\end{align}

Using $\Lambda(\rho_1, u_{\text{th},1}) = (\rho_1/\rho)^2 \Lambda(\rho, u_{\text{th},1})$, one can define the quantity
\begin{align}
    y = \frac{t_* \Lambda(\rho, u_{\text{th},1})}{\rho \left[\beta u_\text{SN} - (1-\beta) u_{\text{th},2}\right]},
\end{align}
a quantity only dependent on the overall density. The cold gas mass fraction $x$ is then given as
\begin{align}
    x = 1 + \frac{1}{2y} - \sqrt{\frac{1}{y} + \frac{1}{4y^2}}
\end{align}

The multi-fluid implementation presents an on-the-fly numerical solution instead of a derivation of an effective equation of state under assumption of equilibrium between the two phases. This implies that additional or more complex source and sink terms are easy to include since it only requires to specify the rates instead of the solution of a complex differential equation. We leave the development of these models to follow-up work and focus in the following on different implementations of the same model to assess the potential and differences inherent to the multi-fluid prescription.

\subsubsection{Star particles and gravitational forces}
Star particles are formed by creating collisionless particles according to a probability derived by the stellar mass formed in a timestep relative to the gas mass of the cell. In the effective equation of state model, a gas cell is stochastically converted to a collisionless particle. In the multifluid model, we preserve the gas cells and keep track of a hydrodynamical and a gravitational mass, the former being continuously drained by star formation, the latter not affected by star formation. Once the difference between the two masses exceeds a certain mass, we spawn a collisinless star particle and reduce the  gravitational mass of the cell accordingly. An important technical detail to note is that the refinement and derefinement keeps the hydrodynamic mass of a cell within a factor of 2 of a specified target mass. This procedure ensures the best possible quality of the hydrodynamics modeling without introduction of additional stochasticity.

For the calculation of the gravitational field, the mass in both phases is simply added up and used as the cell mass. The resulting accelerations are applied as momentum kicks to both phases separately. The calculation and applications of the accelerations are unchanged \citep{springel10, pakmor16}.

\section{Test problems}
\label{sec:verification}

We now show a number of test problems to assess the accuracy of the 2-fluid implementation and its impact on hydrodynamics.

\subsection{Advection problem}
\begin{figure}
    \centering
    \includegraphics{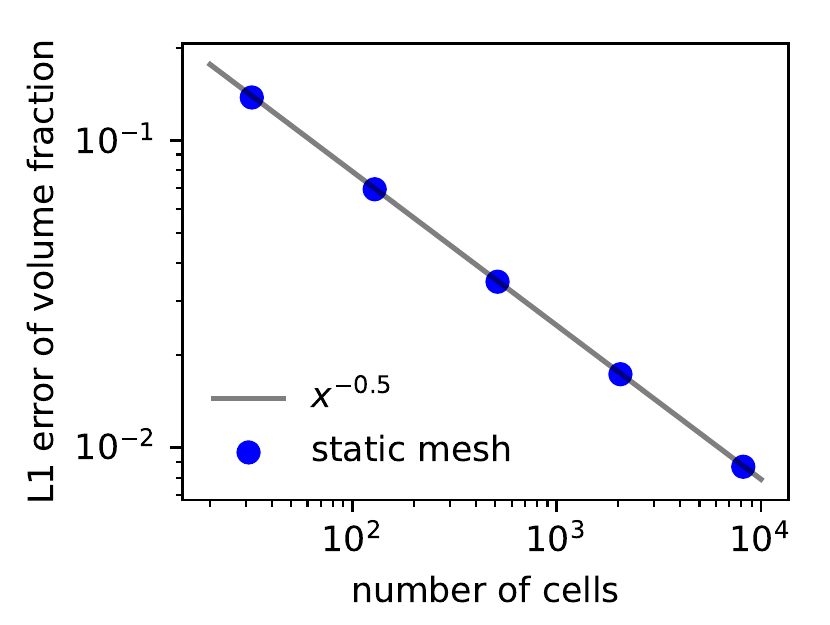}
    \caption{Convergence of the volume fraction in a simple 1d advection test with step-function volume faction variations.}
    \label{fig:convergence_mf_advection_static_mesh_1d}
\end{figure}

\begin{figure}
    \centering
    \includegraphics{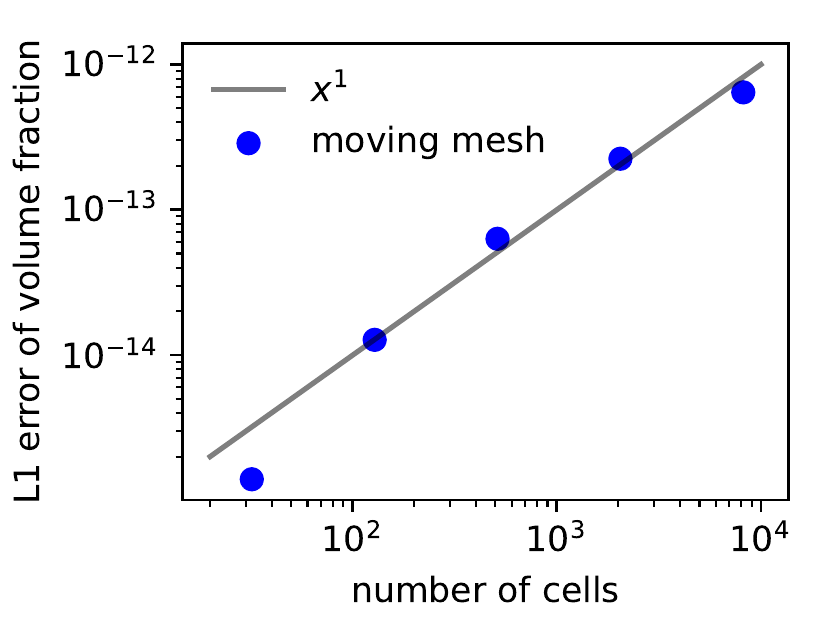}
    \caption{Same as Fig.~\ref{fig:convergence_mf_advection_static_mesh_1d}, only with a moving mesh. The level of error and its scaling with resolution indicate it has origins in floating-point inaccuracies, not discretization errors of the numerical scheme.}
    \label{fig:convergence_mf_advection_1d}
\end{figure}

One of the simplest tests is to set up two fluids with identical density, velocity and specific internal energy, but spatially varying volume fractions. Concretely, we set up $\alpha = 0.25$ for $0.25<x<0.75$ and $\alpha = 0.75$ in the rest, a 1d box with size $1.0$, and periodic boundary conditions. Density, velocity and pressure are $1.0, 1.0$ and $0.6$, the adiabatic index $\gamma=5/3$, the Courant-Friedrichs-Lewy (CFL) factor is $0.3$ in this and all following simulations. This test has two purposes:
\begin{itemize}
 \item In the case of gas at rest, to verify that the code can maintain a pressure equilibrium with varying volume fractions. Formally, this implies in equation~(\ref{eq:momentum}) the last term on the left side to precisely cancel the first term on the right. In our implementation using interface splitting, this is inherently fulfilled, and the code is able to keep the problem static to machine precision (not shown here).
 \item In the case of advection, we test the proper implementation of the time-derivative term in equation~(\ref{eq:energy}) as well as the computation of the volume fraction and pressure at the next timestep. This is verified by making sure that the density and specific thermal energy of the fluids remain constant throughout the simulations (to machine precision), while the respective mass and energy of the cell, i.e., the properties used to compute the density and internal energy of the subsequent timestep, change. To quantify the advection errors of our multifluid scheme, we also compare the volume fractions to the analytic solution after time $t=1$. We show the average L1 error per cell of the volume fraction as a function of the number of cells in Fig.~\ref{fig:convergence_mf_advection_static_mesh_1d}. The advection induced numerical mixing between the two fluids is quite significant and converges only as the square root of the resolution. However, given that our default computational mesh will be moving with the fluid and thereby minimizing the interface fluxes and advection errors, we consider the behaviour acceptable. The convergence in the case of a moving mesh is at machine precision (Fig.~\ref{fig:convergence_mf_advection_1d}).
 \end{itemize}

\subsection{Riemann problem}

\begin{figure}
    \centering
    \includegraphics{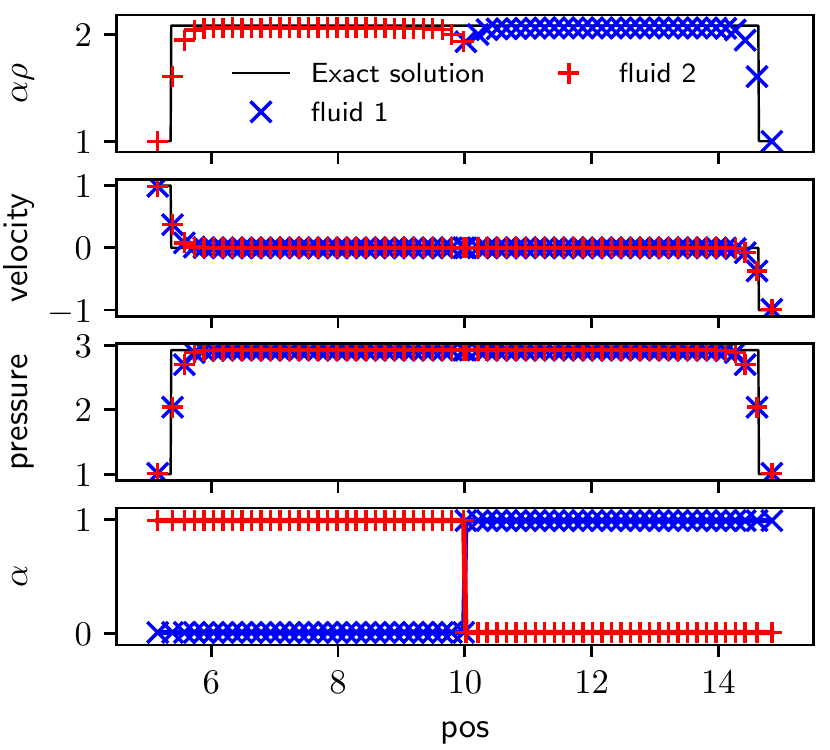}
    \caption{Volume averaged density, velocity pressure and volume fraction of a Riemann problem between two fluids. The black line shows the exact solution. The Riemann problem between different fluid phases is of the same accuracy as the single-fluid version.}
    \label{fig:mf_riemann}
\end{figure}

To test the behaviour of the inter-fluid pressure terms, we set up a symmetric Riemann problem with two converging fluids. Using a 1d box with extent $20$ sampled by $64$ initially uniformly distributed moving cells, we set up the state $\rho = 1.0$, $v(x\leq10) = 1.0$, $v(x>10) = -1.0$, $p=1.0$, for both fluids and the volume fraction of the primary fluid $\alpha(x\leq10) = 10^{-5}$ $\alpha(x>10)=(1-10^{-5})$.  Note that this does not include shear flows between the two fluids which will be discussed later.
Figure~\ref{fig:mf_riemann} shows the result at $t=5$, which turns out to be identical to the single-fluid solution in Arepo (not shown here).

\subsection{Yee vortex}

\begin{figure}
    \centering
    \includegraphics{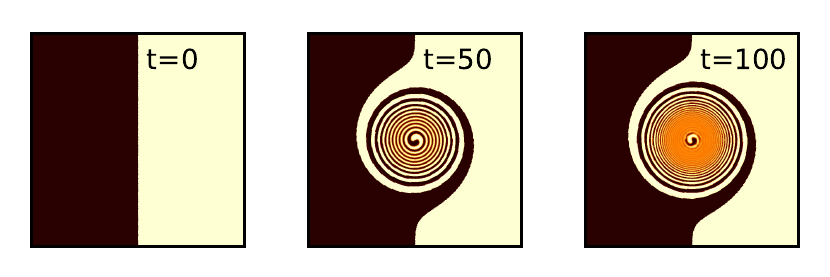}
    \caption{Evolution of the Yee vortex problem with two fluid components. The phases are kept separate until the shear reduces the spatial separation between them to the size of individual cells, at which point the phases numerically mix (orange colour).}
    \label{fig:yee_slice}
\end{figure}

To evaluate the performance of the hydrodynamics in the presence of two fluids, we perform a convergence study with a \citet{yee00} vortex problem with two fluid components.
The simulation setup is identical to \cite{pakmor16} using the implementation presented in \cite{weinberger20}, with the addition that the volume fraction is $\alpha=0.25$ on the right half and $\alpha = 0.75$ on the left half of the box. The simulation time is  $t=100$; i.e. $10$ times longer than the original setup. This is done to follow the spiraling up and numerical mixing of the two phases in the center (see Fig.~\ref{fig:yee_slice}).

\begin{figure}
    \centering
    \includegraphics{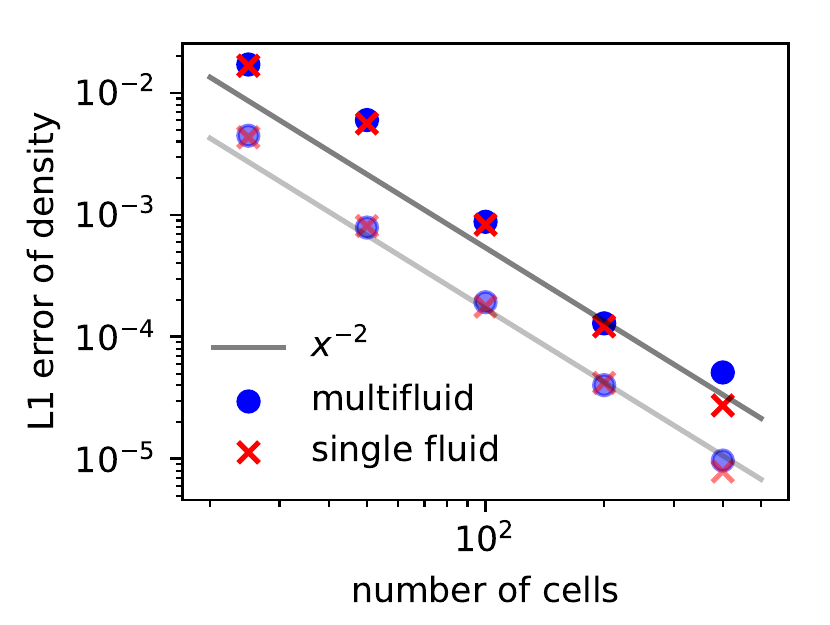}
    \caption{Convergence of the Yee vortex problem with a single and with 2 fluid components. The lower data points are at $t=10$, the upper ones at $t=100$. The respective scaling lines are shifted by a factor of $\sqrt{10}$, for ease of comparison.}
    \label{fig:yee_convergence}
\end{figure}

We show the convergence in density in Fig.~\ref{fig:yee_convergence}, both for the two-fluid setup as well as the original single-fluid problem. The lower points are at $t=10$, the upper ones at $t=100$. This shows not only that the two-fluid approach preserves the second order convergence of Arepo, but also with the same normalisation as in the single-fluid case. In short, the presence of a second fluid does not affect the quality of the hydrodynamics.

\subsection{Source terms: a two-fluid model for the ISM}

\begin{figure*}
    \centering
    \includegraphics{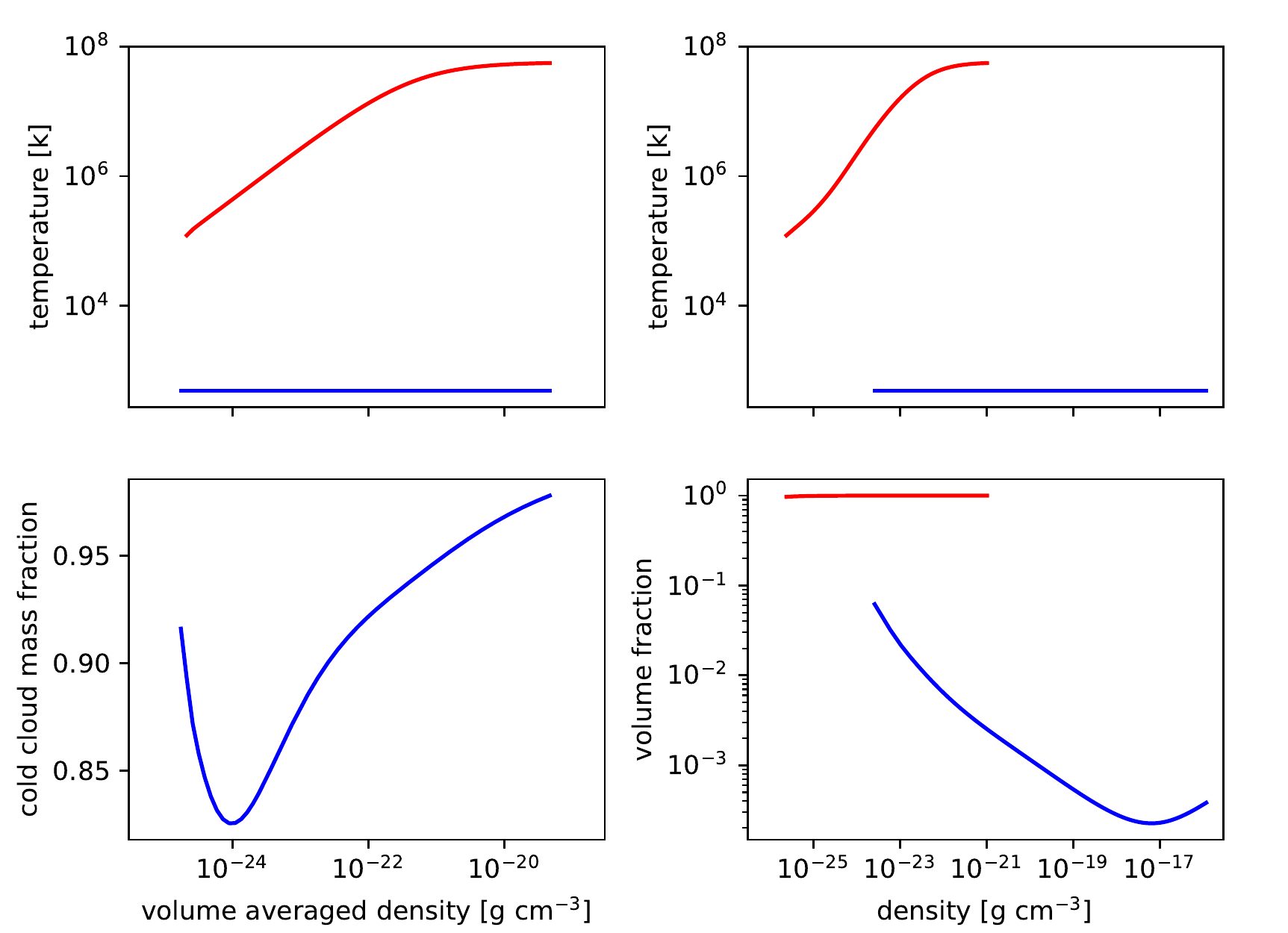}
    \caption{Equilibrium state due to source terms in absence of resolved hydrodynamic fluxes of hot (red, phase 1) and cold (blue, phase 2) gas cells in 2 phase ISM model.}
    \label{fig:phases_ism}
\end{figure*}

To test the ISM model, we set up gas with uniform density and add a mass source term to gradually increase the density, however, slowly enough to allow inter-phase terms to reach an equilibrium state. This way we can study how the multi-phase properties of the gas behave as a function of density due to source terms. The result is shown in Fig.~\ref{fig:phases_ism}, where we show temperature and mass fraction of the cold phase vs. volume averaged density in the left two panels and temperature and volume fraction vs. physical density in the right panel. The left panels can be directly compared to \citet{springel03}; the right panels are more relevant to the multi-fluid hydrodynamics equations. The first thing to notice is the substantially higher densities reached by the cold gas, due to the isothermal nature of the cold gas at $1000$~K and the pressure-equilibrium assumption. The second is the fact that while the mass fraction of the cold phase is substantial, the volume fraction is always very small. This particularly means that the pressure terms of the fluxes for the cold phase will be small, while the mass in this phase is dominant. This, in turn also leads to a different behaviour in the hot phase compared to the effective equation of state implementation of \citet{springel03}, which is at the core of why we use this technique in the first place.

\begin{figure}
    \centering
    \includegraphics{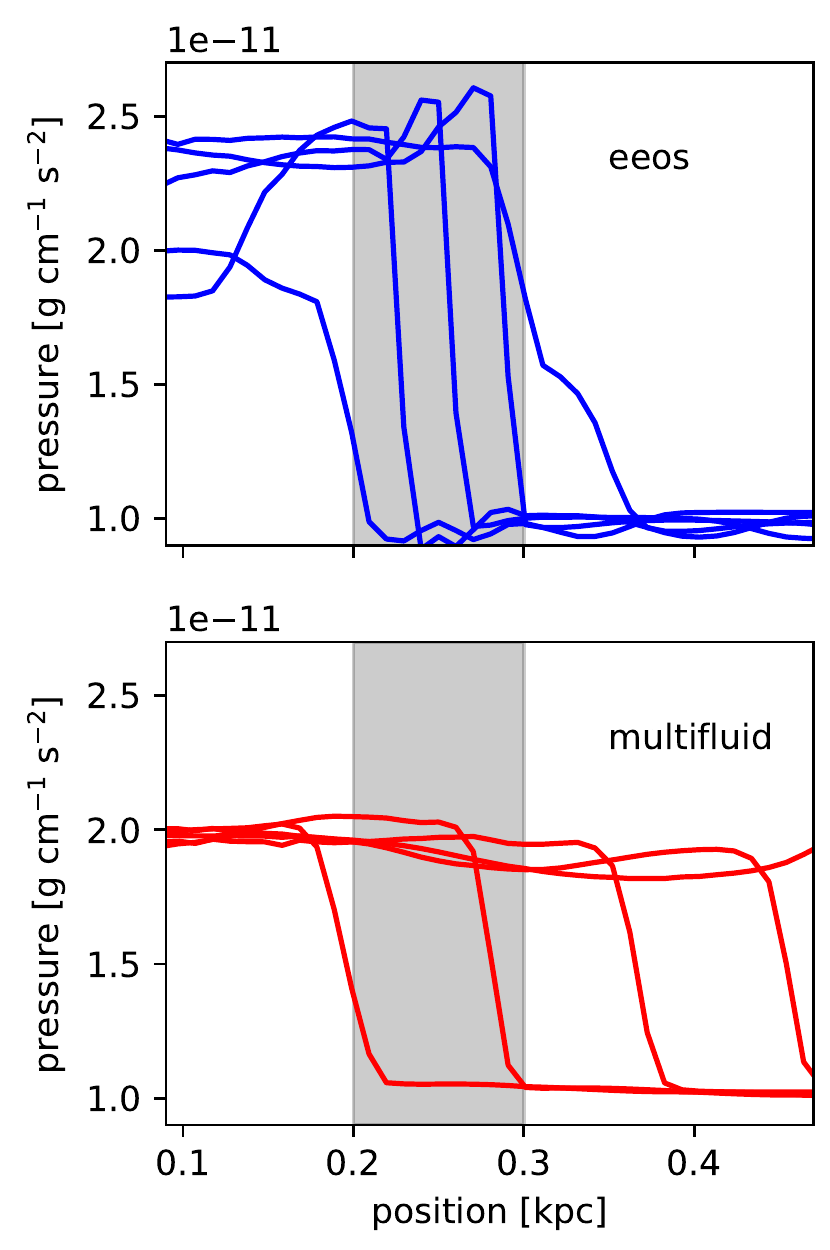}
    \caption{Pressure vs. position in x-direction for the shock test though a narrow column of 2-phase interstellar medium (grey shaded area) embedded in a background medium in pressure equilibrium. The different lines show the pressure at different times (from left to right: $0.5$, $1.0$, $1.5$, $2.0$, $2.5$~Myr). The effective equation of state causes the shock to be partially reflected and to propagate slower, while the 2-fluid model allows an almost unperturbed penetration though the ISM medium at the shock propagation speed of the hot phase.}
    \label{fig:shockthroughism}
\end{figure}
To illustrate this, we set up a box with side length $0.5$~kpc, $50^3$ cells and open boundary conditions. The background density of the gas is set to $9.8\times10^{-26}$~g~cm$^{-3}$, below the star formation threshold of the interstellar medium model (ISM source terms / effective equation of state inactive), and the background temperature to $7.7\times 10^5$~K assuming a mean particle mass $\mu=0.6$ times the hydrogen mass. In cells with x-coordinate in the range $[0.2 \,\text{kpc},0.3 \,\text{kpc}]$, we additionally add a cold phase with mass fraction $0.9$ and temperature $1000$~K, leading to this strip being 2-phase with ISM source terms active (or the effective equation of state in the comparison simulation), in approximate pressure equilibrium with the background. The volume fraction of the second fluid everywhere else in vanishing ($10^{-12}$). We subsequently increase the temperature of the cells with x-coordinate $<0.1$~kpc by a factor of $3$ to induce a shock which propagates through the ISM region. As the volume fraction of the cold phase is small, we expect the propagation of the shock to not be overly impacted by the presence of the cold phase. The source terms could in principle absorb some of the shock energy from the hot into the cold phase, however the timescale of source-term mass exchange is significantly larger than the shock crossing time. We therefore expect only a small damping of the shock. The evolution of the pressure (every $0.5$~Myr) is shown in Fig.~(\ref{fig:shockthroughism}). The 2-fluid approach shows exactly the expected behaviour. The effective equation of state model, however, slows down the shock propagation in the 2-phase area, causing a partial reflection and a damping of the forward shock. This behaviour is a numerical artifact caused by a locking of the two phases and illustrates the advantages of the 2-fluid prescription even when using the identical source term model. While this example is deliberately simple, shocks propagating though the interstellar medium are common, both externally driven (e.g., ram-pressure stripping, galaxy mergers, etc.) or from within the galaxy (e.g., feedback from active galactic nuclei) making these improvements important in certain conditions. We leave the study of these cases for future work and focus on the application to various more general galaxy formation simulations in the following.

\section{Application}
\label{sec:application}

We apply the multifluid implementation combined with the source terms of the two-fluid ISM model in frequently used galaxy formation simulations.
We compare the results to the respective simulations applying an effective equation of state assuming the same model.
Broadly, we expect very similar but not necessarily identical results:
the main difference between the two models is the ability for non-equilibrium ISM states, as well as relative velocities of the two phases in the multifluid implementation.
This implies that we expect the largest deviation between the two solutions during rapidly changing, dynamical environments such as galaxy mergers.

\subsection{Disk galaxy}

\begin{figure*}
    \centering
    \includegraphics{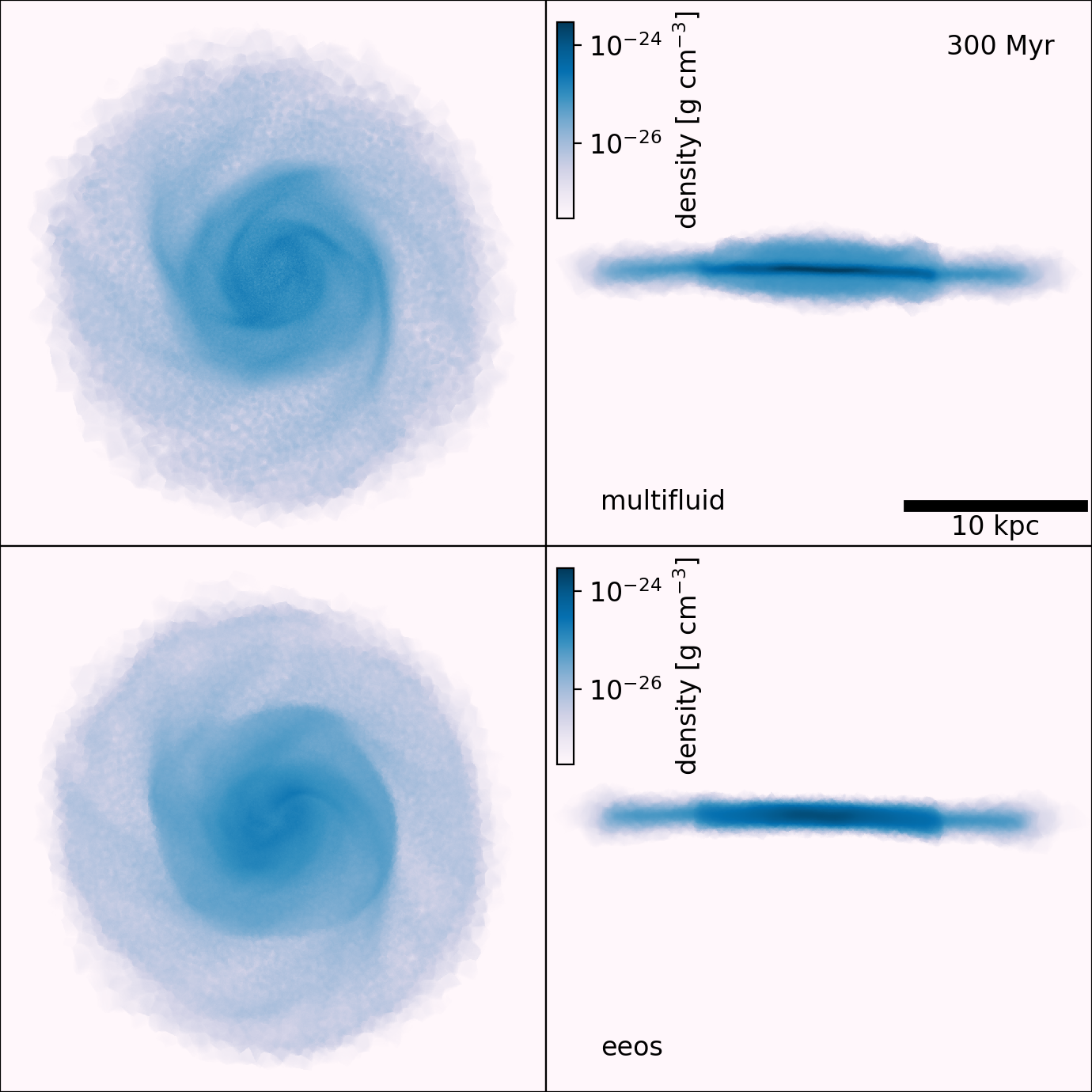}
    \caption{Total mass density (of both phases)  face-on and edge-on in isolated disk galaxies volume averaged over $10$~kpc depth. The upper plots represent the run with the 2 fluid model, the lower plots with the traditional effective equation of state model.}
    \label{fig:disk_galaxy}
\end{figure*}

\begin{figure}
    \centering
    \includegraphics{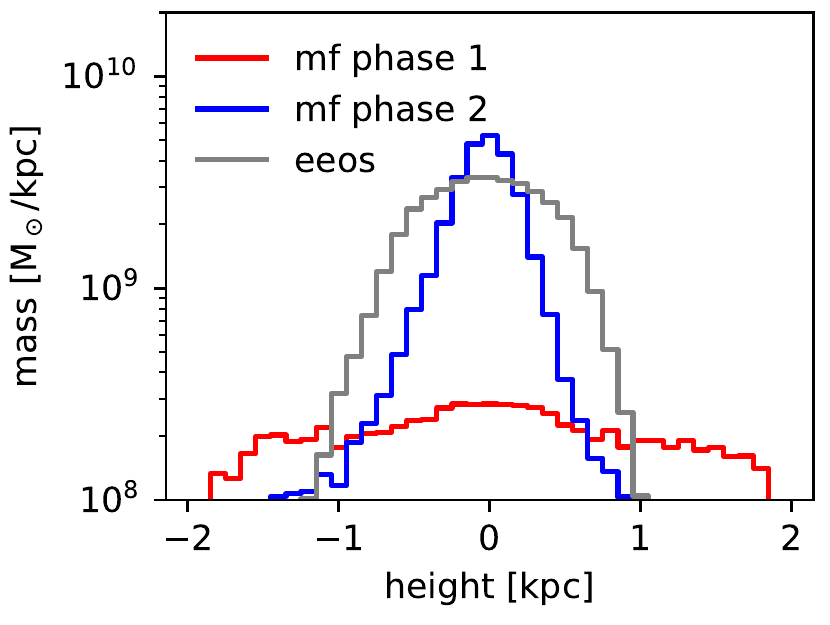}
    \caption{Mass distribution of different phases in isolated disk galaxy simulation using the 2-fluid model at time $t=300$~Myr, restricted to gas in the inner $10$~kpc. The cold phase 2 clearly shows a smaller scaleheight while the hot phase 1 is more vertically extended. The grey line shows the effective equation of state disk for comparison.}
    \label{fig:disk_galaxy_scaleheight}
\end{figure}

We start out using a single disk galaxy in equilibrium from the `galaxy merger' example in \citet{weinberger20}. This disk galaxy is in equilibrium and secularly evolving over a little more than $1$~Gyr, after which its gas supply is substantially depleted. The gas density is shown in Fig.~\ref{fig:disk_galaxy}, face-on and edge-on for both multi-phase (top) and effective equation of state (bottom) runs. As expected, the two modeling techniques produce similar results, but there are a few notable differences: the inner part of the effective equation of state model maintains a slightly puffed-up dense disk without any material at larger heights. This is a direct consequence of the pressure dictated by the effective equation of state. The initial conditions are set up such that vertical hydrostatic equilibrium is ensured, thus no  material is launched from the disk. The multi-phase model, however, behaves slightly differently: a thinner, denser disk is visible at the radii where the ISM model is active, surrounded by an even thicker, dilute disk with large scale-height. This trend is shown quantitatively in Fig.~\ref{fig:disk_galaxy_scaleheight}. The time evolution of these simulations reveals that the dense, cold second fluid sediments towards the center, causing an increase in density there. This non-equilibrium process triggers some of the hot gas to be lifted upward, forming a more puffy disk. Note however that the setup subsequently reaches an equilibrium state; i.e. the dilute thicker disk does not form an outflow. For this model to create an outflow, the source terms would need to trigger non-equilibrium states in pressure, an extension we leave to future work.

\begin{figure}
    \centering
    \includegraphics{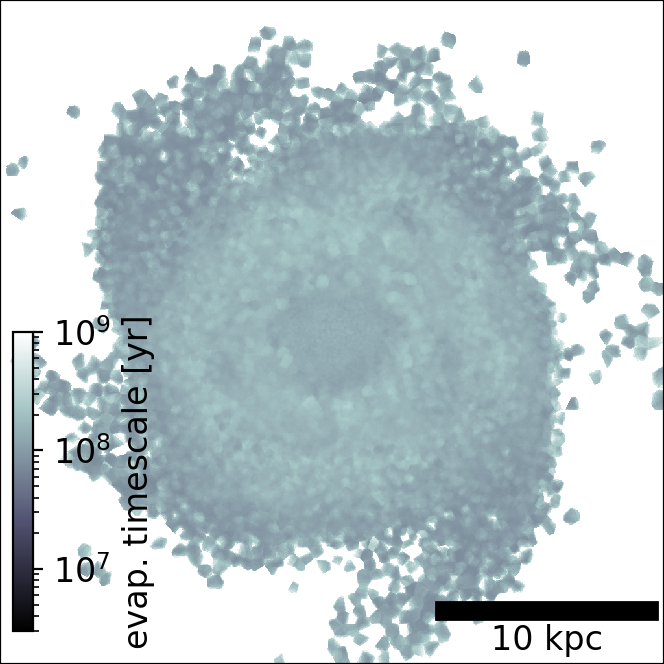}
    \caption{Evaporation timescale of the cold phase due to supernova. Note that the timescale is around $10^8$~yr, slightly smaller than the orbital timescales, underscoring the importance of the mass exchange between phases.}
    \label{fig:disk_galaxy_evaporation}
\end{figure}

Qualitatively, the thinner cold gas disk is an expected outcome caused by the decoupling of the velocities of the hot and cold phases, thereby allowing the sedimentation of the cold gas clouds. Quantitatively, Fig.~\ref{fig:disk_galaxy_scaleheight} shows a decrease in scaleheight by a factor of about 2, less than expected from the decrease in pressure support. While in a real galaxy, such a behaviour can be explained by the development of a vertical velocity dispersion between the clouds, it is important to realize that our 2 fluid model does not allow such a velocity dispersion below the resolution scale (due to the assumption of a single velocity for phase 2 per resolution element). The origin of this cold, vertically non-pressure-supported disk in the simulation becomes clear when considering the mass exchange terms between the phases, in particular the evaporation timescale of the cold cloud due to supernovae. Fig.~\ref{fig:disk_galaxy_evaporation} shows a map of this timescale, indicating that it is of order $100$~Myr, implying a substantial mass (and consequently momentum) exchange between the phases. This means that for secularly evolving disk galaxies, hot and cold phases, while not exactly cospatial, cannot be considered decoupled. Aside from the scaleheight, this coupling also acts as a stabilizing effect against gas clumping in the disk.

\begin{figure}
    \centering
    \includegraphics{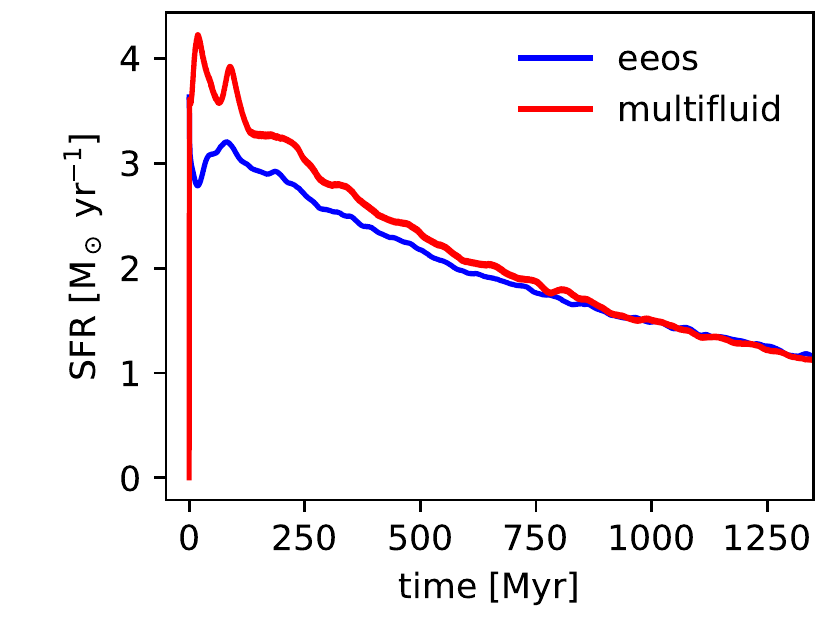}
    \caption{Star formation rate of disk galaxy simulated with multifluid and effective equation of state star formation models.}
    \label{fig:disk_galaxy_sfr}
\end{figure}

Figure~\ref{fig:disk_galaxy_sfr} shows the star formation rate of the secularly evolving disk galaxy. Consistent with the notion of the multi-phase simulations being slightly out of equilibrium initially, the star formation rate in the first $500$~Myr is slightly higher, and oscillating slightly. After 300~Myr (the time shown in Fig.~\ref{fig:disk_galaxy}) both galaxies have very similar star formation rates. Overall, we conclude that the two models yield very similar results with differences that can be understood due to the differences in the assumptions made in the equations that are solved.

\subsection{Galaxy merger}

\begin{figure}
    \centering
    \includegraphics{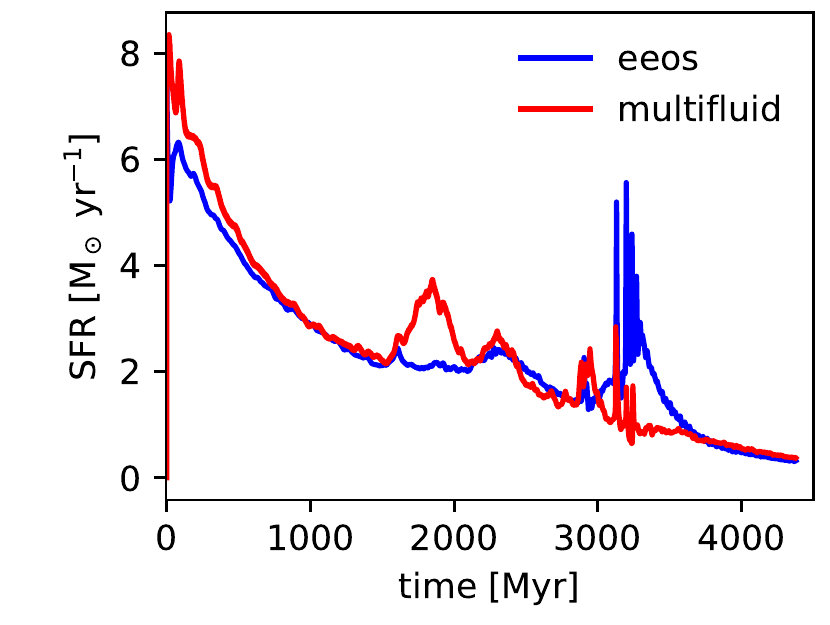}
    \caption{Star formation rate of two merging disk galaxies with multifluid and effective equation of state star formation models.}
    \label{fig:galaxy_merger_sfr}
\end{figure}

Having shown that the secular evolution is well understood, we test the merger of two disk galaxies; i.e. reproducing the \citet{weinberger20} galaxy merger example and using the multi-phase model for the same setup. The resulting star formation rates are shown in Fig.~\ref{fig:galaxy_merger_sfr}. The differences in star formation rate are somewhat larger than in the case of secular evolution, however, for the most part they are still similar. The largest differences in star formation rate occurring during tidally induced starbursts at first and second pericenter as well as at final coalescence around $3500$~Myr. This is the time when the gas properties in the star-forming galactic center change rapidly, which likely explains the larger deviation of the two models and underscores the importance of replacing the equilibrium assumptions of the effective equation of state model with an explicit time integration of the underlying rate equations when studying these rapidly changing galactic environments.

\subsection{Cosmological volume with cooling and star formation}

We now move on to cosmological simulations. These have become increasingly important due to their ability to model a statistically complete sample of galaxies from well-constrained initial conditions, yet suffer from resolution limitations due to limited computing and memory resources.
These types of simulations would particularly benefit from an accurate modeling of an unresolved multi-phase medium.
In this section, we show only a few relatively inexpensive examples of a very small volume to illustrate the model's behaviour and leave runs covering larger volumes to future work.
The initial conditions are created using the MUSIC code \citep{hahn11}, with the parameters of the cosmological volume example of \cite{weinberger20}. However, we simulate the volume at higher resolution with up to $2\times128^3$ resolution elements in a volume of comoving side length $7.5\, h^{-1}$~Mpc and accordingly lower softening of $1.5\, h^{-1}$~kpc (comoving) but at most $0.75\, h^{-1}$~kpc in proper coordinates. The respective mass per gas cell/dark matter particle is $2.7\times 10^6\, h^{-1}$ and $1.5\times10^7\, h^{-1}$~M$_\odot$, respectively, similar to state of the art large volume cosmological simulations.

\begin{figure}
    \centering
    \includegraphics{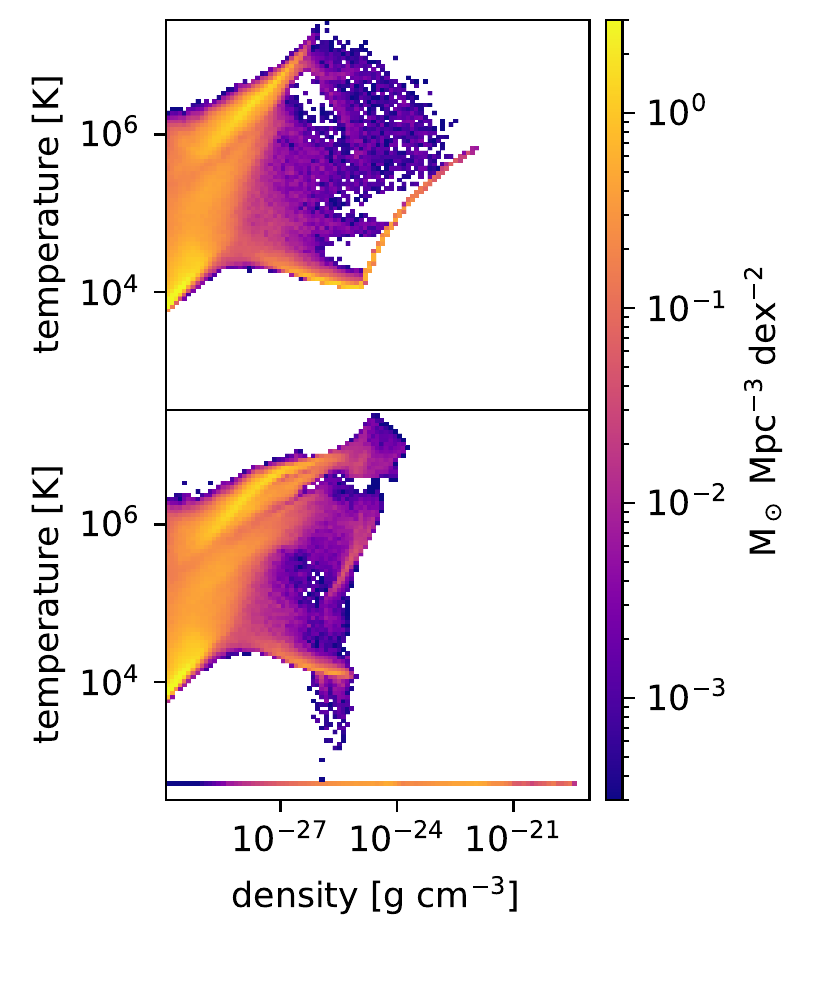}
    \caption{Phase diagram of gas in the cosmological volume simulation using the effective equation of state model (top) and the multifluid implementation (bottom). The key difference is the explicit presence of cold, isothermal gas in the multifluid implementation reaching higher densities than the effective equation of state model, as well as more hot gas at lower densities. The vast differences are simply a consequence of the different definition of density, with the one in the effective equation of state being a large-scale volume average while in the multifluid model it refers to the density in the respective phase; however, in both cases, this is the input into the hydrodynamics solver.}
    \label{fig:phase_diagram_cosmo_box}
\end{figure}

In Fig.~\ref{fig:phase_diagram_cosmo_box}, we show the phase diagram of the gas in the simulation at redshift $z=0$. As expected, the low-density, low-temperature regime appears very similar between the simulations, as gas in this state is mostly non-radiative and single-phase. The high temperature gas differs a bit, with the multifluid model having more of it exceeding $10^{-25}$~g~cm$^{-3}$. This gas is the hot component of the star forming cells, which in the effective equation of state model makes up the characteristic high density tail of the phase diagram. The multifluid model has (by construction) its second phase at $1000$~K. The densities are then given by the pressure equilibrium criterion, and thus span the entire range from $10^{-28}$~g~cm$^{-3}$ to beyond $10^{-20}$~g~cm$^{-3}$, with most of the mass in densities above the star formation threshold of a volume-averaged density of around $10^{-25}$~g~cm$^{-3}$. The gas in the primary phase colder than $10^4$~K is likely gas that just barely entered the star forming phase, with a sudden condensation of gas from hot to cold phase creating this feature.

\begin{figure}
    \centering
    \includegraphics{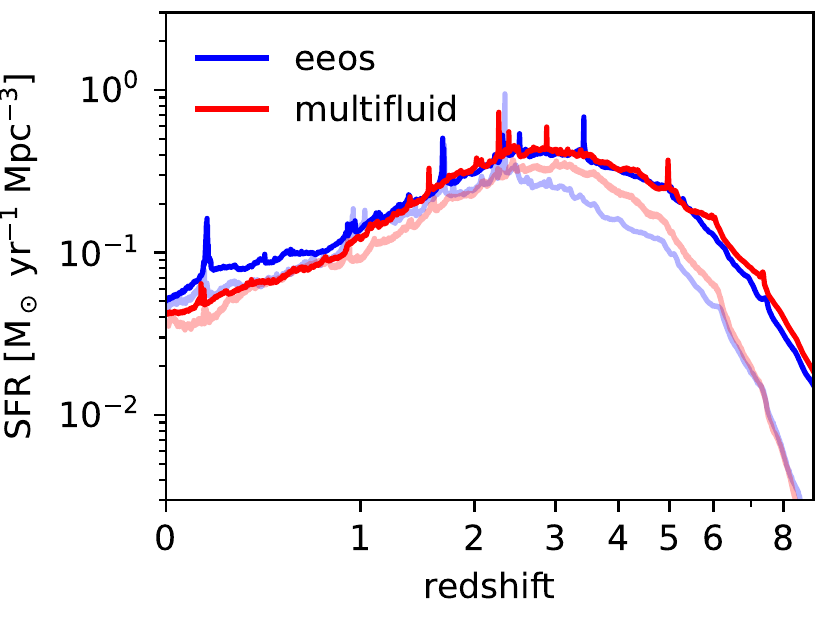}
    \caption{Star formation rate density as a function of redshift in a cosmological volume simulation for effective equation of state and multifluid models. The two implementations exhibit similar behaviour. The transparent lines show lower resolution counterparts of the respective simulations. The spikes in star formation rate density are starbursts of individual systems, i.e. a consequence of the small simulation volume.}
    \label{fig:sfrd}
\end{figure}

The star formation rate densities of the cosmological volumes are shown in Fig.~\ref{fig:sfrd}, for the reference run using the effective equation of state model as well as for the multifluid implementation. The high resolution versions (thick lines) show very similar star formation rate densities between the models. For the lower resolution counterparts (thinner lines, $2\times64^3$ resolution elements), the models differ more clearly, however not as much as the different resolutions. We thus conclude that our implementation yields the a similar result also in cosmological simulations.

\subsection{Cosmological zoom simulation}

\begin{figure}
    \centering
    \includegraphics{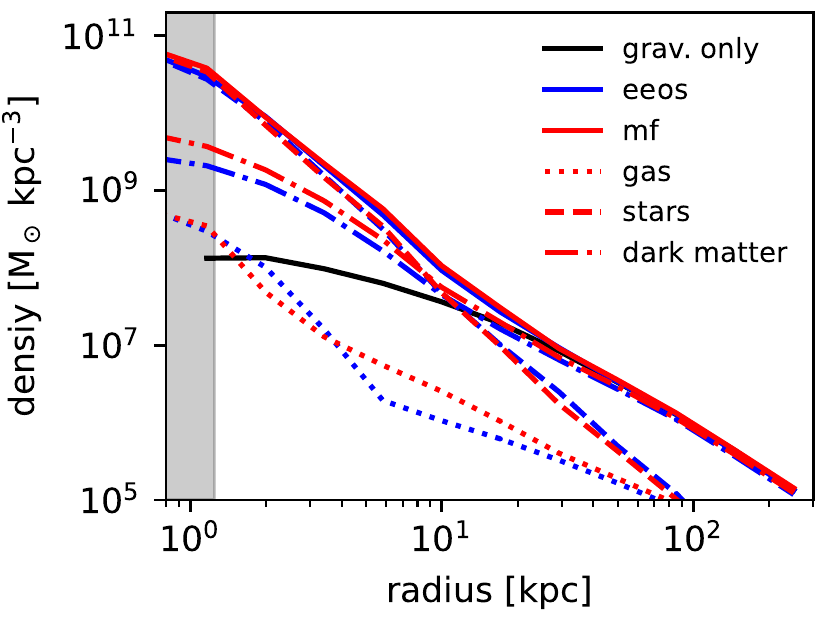}
    \caption{Mass profiles of central halo of cosmological zoom simulation, run as a gravity only version (black), with effective equation of state model (blue) and multifluid model (red). Dashed, dash-dotted and dotted lies denote stellar, dark matter and gas density, respectively, and the solid line the total density profile. The multifluid and effective equation of state models show a distinct steepening of the inner density profile relative to the gravity only version. The shaded area denotes the gravitational softening length.}
    \label{fig:zoom_profiles}
\end{figure}

Along with cosmological volumes with uniform resolution, cosmological zoom simulations play an important role in galaxy formation research. The zoom technique makes it possible to focus on the cosmological formation of an individual object, thereby allowing a substantially higher resolution at fixed computational resources. Here, we use the demonstration example of a moderately massive galaxy cluster presented in \cite{weinberger20} at slightly higher resolution. The initial conditions are created using the MUSIC code \citep{hahn11} in a cubic volume of length $100$~Mpc~$h^{-1}$, with a maximum level of refinement set to level 10; i.e. the box is divided by $2^{10} = 1024$ elements at the highest resolution grid in the initial conditions, resulting in a mass per resolution element of $8.0\times10^7$~M$_\odot\, h^{-1}$. These initial conditions are then split into collisionless (i.e. dark matter) and gas components of mass $6.7\times10^7$~M$_\odot\, h^{-1}$ and $1.3\times 10^{7}$~M$_\odot\, h^{-1}$, respectively. The initial volume fraction of phase 2, i.e. the cold gas, is  $10^{-8}$.

Fig.~\ref{fig:zoom_profiles} shows the radial density profiles of the main halo at redshift~0 for 3 different simulations: one with the multifluid model (red), one using the effective equation of state model (blue), and a gravity only simulation (black) for reference. For the former two models, the dotted line indicates the gas density (both phases), the dash-dotted lines the dark matter density and the dashed line the stellar density profile. The solid lines show the total density. The similarity between both models with star formation indicates that the gravitational interaction between the phases is captured accurately and stars form not only at the same rate but in the same locations for both models. Overall, we have thus shown that the 2-fluid gas modeling yields very similar results in cosmological simulations than the corresponding effective equation of state modeling. Combined with the increased flexibility and more straightforward extension to more complex interaction terms it represents an ideal basis for future development of more complex models of the ISM in these types of simulations.

\subsection{Multi-phase winds}

\begin{figure*}
    \centering
    \includegraphics{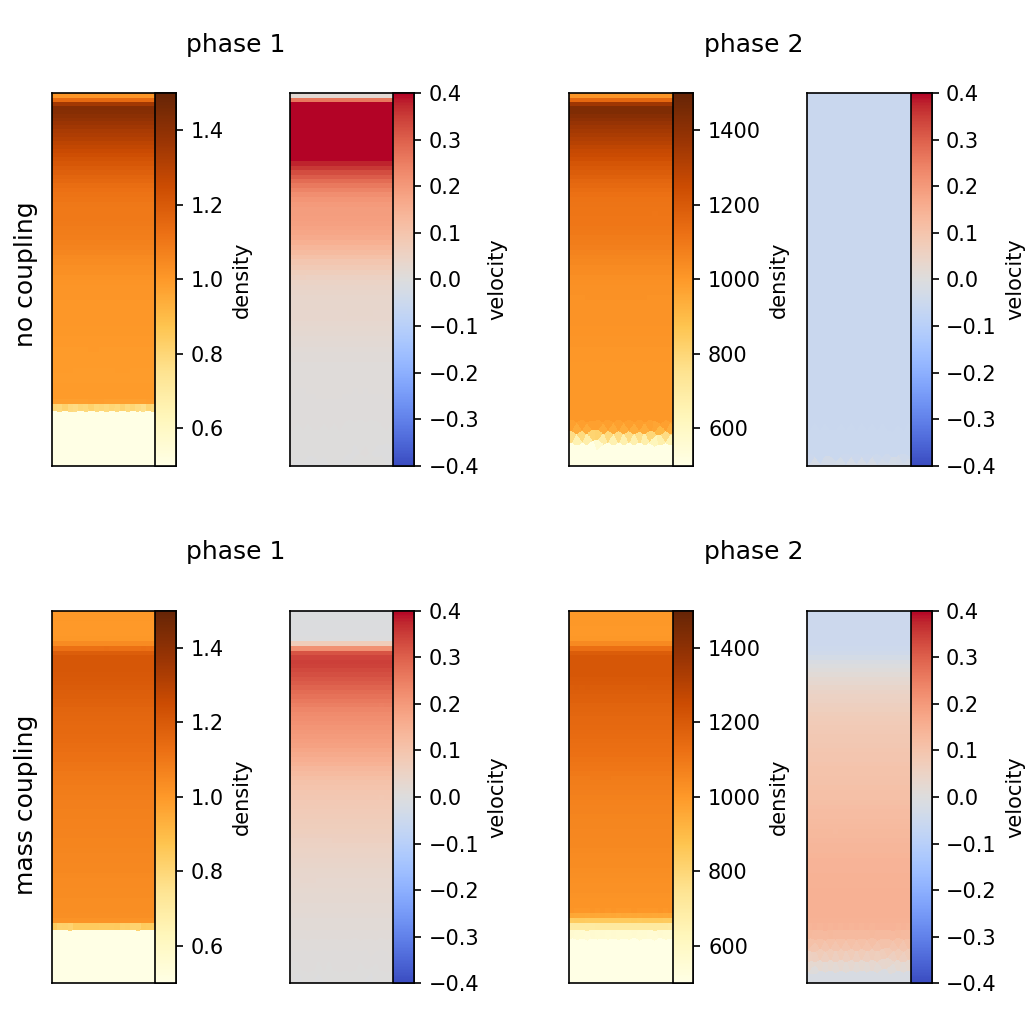}
    \caption{Density and outflow velocity of hot (left) and cold (right) phases. The upper plots show the simulation without source terms, the lower a momentum coupling via mass exchange between the phases. The cold phase is accelerated mostly by a mass-exchange term with the hot phase, while a shock runs though the hot phase.}
    \label{fig:outflow}
\end{figure*}

A further case of interest for galaxy evolution is a stratified medium, modeling the vertical patch of a disk galaxy in order to study the ISM and stellar-feedback driven galactic outflows.
These simulations have been frequently applied to understand the low-altitude multi-phase nature of the outflows \citep{walch15, kim17b} and their dependence on driving mechanisms \citep{simpson16, girichidis18}. Here, we only show a simplified version of this setup: a simple stratified medium, overpressurized at the bottom, driving a plane-parallel shock outwards. We use a 2d computational box $0.5$ times $2.5$ and a constant gravitational pull in the negative y direction $g_y=-0.1$. The density in both phases $\rho_1$, $\rho_2$, volume fraction of phase 1 $\alpha$, velocities $\vec{v}_1$, $\vec{v}_2$ and pressure $p$ in the initial conditions are
\begin{align}
    \rho_1 &= 1.0 \nonumber\\
    \rho_2 &= 1000.0 \nonumber\\
    \alpha &= 1.0 - 10^{-3} \nonumber \\
    \vec{v}_1 &= \vec{v}_2 = 0 \\
    p &= 2.5 + g_y\, (y-1.25) \,\rho_1 \nonumber \\
    p&(y<0.1) = 3.0 \left(2.5 + g_y\, (y-1.25) \,\rho_1\right). \nonumber
\end{align}

The subsequent evolution at time $t=0.54$ is shown in the upper panels of Fig~\ref{fig:outflow}. The left two frames show the density and velocity of phase~1 and the right two panels show the same quantities for phase~2. The plane parallel shock is reaching the upper end of the box at this point, and the resulting compression can be seen in the density slice in both phases (note that the density of the second phase is 3 orders of magnitude higher). However, the upward velocity this shock induces occurs only in phase 1. This is expected, since the volume fraction of phase 2 is $10^{-3}$ and consequently pressure gradients do not lead to the same outward acceleration. Indeed, we expect a sedimentation of the cold gas in spite of the shock traveling outwards close to $v_y(t)\approx g_y\, t = -0.054$, which is indeed what can be seen in the top right panel of Fig.~\ref{fig:outflow}.

Galactic outflows are observed to be multi-phase \citep[see][for a review]{veilleux20}. As previously discussed, accelerating a cold phase via pressure gradients does not work in practice, so other ways of producing a cold outflowing phase need to be considered. Two basic scenarios are possible: either drag forces lift up cold gas or mass condensates out of the hot phase. Within our numerical model, both processes would show up as source terms. In this particular setup, we assume a condensation from phase 1 onto phase 2 at the rate
\begin{align}
    \dot{m} &= (1-\alpha)\, \rho_2\, V\, t_\text{cond.}^{-1} \\
    t_\text{cond.} &= 0.1 \left|\vec{v}_1-\vec{v}_2\right|^{-1}
\end{align}
and furthermore assume the evaporation rate from phase 2 to phase 1 to be equal (note that there is no physical reason for this, but it simplifies the example considerably). This implies that there is momentum exchange between the phases proportional to their mass exchange rate times the relative velocity, while the mass in each phase remains constant. The simulation result with this mass exchange term is shown in the lower panels of Fig.~\ref{fig:outflow}. Two aspects are noteworthy here: first, phase 2 develops an outflow, trailing phase 1 and not quite reaching the same velocities. Second, the position of the shock is lower compared to the non-coupling simulation. This is understandable since the exchange of momentum implies an energy loss of the shock-driving phase 1.

We note that while the simulation presented here was run with $24$ times $120$ resolution elements, a qualitatively similar result can be obtained with far coarser resolution due to the multi-phase model being agnostic to the precise spatial distribution and size of the cold gas clouds. This means that the employed source terms, i.e. the coupling between the phases, can be parameterized and calibrated to dedicated high-resolution outflow studies, implemented into the multiphase model and verified at resolutions of cosmological simulations (using the presented outflow setup) and then used confidently in cosmological simulations to study the impact on galaxy populations.

\section{Discussion}
\label{sec:discussion}

We propose a 2-fluid finite-volume discretization to model multi-phase gas in galaxy scale numerical simulations. We implemented the model into the moving-mesh finite-volume code Arepo, and show:

\begin{itemize}
    \item Our multi-fluid solver is able to maintain interface discontinuities between different phases in pressure equilibrium.
    \item While the advection errors in volume fraction are substantial, this shortcoming is in practice not the dominant concern when using a mesh moving with the fluid flow (the default mode of operation in Arepo).
    \item The performance of the hydrodynamics solver is practically unaffected by the presence of a second fluid component, both in terms of convergence for smooth problems and for shocks.
    \item Existing multi-phase ISM models can be implemented by solving the underlying differential equations on the fly. This approach implies that additional source and sink terms can be added trivially.
    \item Unlike effective equation of state models, the multi-phase gas can be modeled without 'locking' the hot phase (both kinematically or thermally). This means that there is no need for a hydrodynamical decoupling, a delay of cooling or an unphysical accumulation of energy in feedback models; i.e. it solves the numerical overcooling problem in the ISM of cosmological simulations.
    \item Commonly used simulation setups such as an isolated disk galaxy, a galaxy merger, cosmological volume simulations and cosmological zoom simulations show very similar behaviour comparing the multi-fluid and effective equation of state approaches, with expected differences in specific situations. This is done as a verification of the implementation. We note however that there is no explicit feedback in the simulations; i.e. the presented runs are in this sense considerably simpler than modern cosmological volume simulations. We speculate that the differences between the treatments would be larger for models including explicit feedback.
    \item Simulations of a shock moving outwards though a stratified medium shows how source terms can be used to model the driving of multi-phase outflows via condensation and evaporation without the need to resolve the clouds making up the cold phase. Driving via drag forces can be handled in the same way.
\end{itemize}

The 2-fluid modeling thereby also improves the prospects of more accurately connecting resolved with unresolved scales:
\begin{itemize}
    \item The model can be run in cosmological simulations. The production runs in this work are computationally a bit more expensive, however, mostly due to timestep constraints in the explicit integration of source terms. Switching to an implicit time-integration for source terms should in principle remove this constraint and reduce the difference in computational resources needed to a factor of a few, vastly outperforming attempts to spatially resolve the cold phase both in computational and memory requirements.
    \item Small scale effects are parameterized through source terms that are mathematically well-defined by the discretization procedure itself. To coarse-grain a small-scale simulation, it is clear which properties need to be measured and parameterized as a function of large-scale hydrodynamic quantities.
    \item The coarse grained model can be straightforwardly verified by simply repeating the small-scale simulation at lower resolution and comparing results.
    \item We showed the example of a multi-phase outflow, but this is also possible in other cases such as thermal instability in the ICM and shear flows between hot and cold media.
\end{itemize}

Overall, we believe that this kind of modeling is able to overcome some of the most restrictive limitations plaguing galaxy formation simulations to date, yet more detailed work needs to be done to conclusively show this in each individual case.

\section{Conclusions}
\label{sec:conclusions}

We present an implementation of a compressible 2-fluid model in the Arepo code \citep{springel10} using a finite-volume moving mesh approach and a stratified flow model \citep{chang07}. This discretization is able to handle both unresolved mixtures as well as resolved phase boundaries. We show that while advection of such phase boundaries across a grid is fairly diffusive and only converges with the square root of the number of cells, the moving mesh technique alleviates these problems in practice. The accuracy and convergence properties of hydrodynamics test problems are unchanged by the presence of a second fluid both for shock tests and smooth problems such as a \citet{yee00} vortex.

We explore the use of this model in a two-phase interstellar medium modeling with \citet{springel03} source terms and compare the performance in typical simulations of galaxy formation and evolution: isolated galaxies, idealized galaxy mergers, cosmological volumes with uniform resolution and cosmological zoom simulations and show that results are similar to the established effective equation of state model, only differing in some details in predictable ways. Applied to a simplified test resembling galactic winds, we demonstrate the ability to use source-terms to model mass exchange and the driving of multi-phase outflows.

Our results suggest that the presented model is a viable alternative to the effective equation of state \citep{springel03, schaye08, dubois08, dave16} or resolved low-temperature cooling modeling in cosmological simulations \citep{hopkins12, agertz13, smith18, marinacci19}, while also alleviating some of its most restrictive limits. Most notably:
\begin{itemize}
    \item Cold and hot phases are not locked, porosity on unresolved scales is not an issue.
    \item A clear and easy way to implement complex small scale effects via source/sink terms.
    \item The numerical timestep is not tied to the smallest physical structures, decreasing computational cost of the simulations compared to explicitly resolved simulations.
\end{itemize}
This is done while maintaining the clear separation between resolved and unresolved scales, making future coarse-graining efforts of high resolution studies of specific effects a promising avenue.
As such, the multi-fluid technique addresses some of the limiting factors of present-day cosmological simulations and has the potential to be transformative for our ability to model the interstellar, circumgalactic and intra-cluster gas with moderate computational resources, as well as the ability to study the effect of complex small scale physics in a cosmological context in ways that were not previously possible.

This work, however, only represents the starting point towards this goal. A generalized model for source terms needs to be developed, applying not only to dense, but to all gas. In the ISM case, the employed equilibrium model needs to be expanded to more realistically reflect the state of the ISM \citep[see e.g.][]{buck22}. Most importantly, it needs to be able to self-consistently drive galactic winds (which require non-equilibrium states in the ISM), possibly via the introduction of stochastic terms. For the CGM and ICM, ensemble models for condensation, interaction and evaporation of cold clouds (and their dependence on environment) need to be developed.

On the numerics side, an implicit integration of the source terms needs to be explored to avoid small timesteps due to stiff source terms which at present limit the performance. Further exploration of the stability of different pressure-closures might reveal more suitable multifluid models. Higher order reconstruction of the volume fractions might improve the quality of the volume-fraction advection problem. Further pressureless fluid components can be implemented, e.g. to model partially coupled dust which can include sophisticated formation and destruction terms, as well as coupling to the fluid and absorption/emission of radiation. Finally, a mass diffusion term of the cold gas component could be implemented to model the intrinsic velocity dispersion of unresolved cold gas clouds in a statistical manner.

In summary, this work represents a proof of concept for a new way of modeling multi-phase gas in numerical simulations of galaxy formation. Future work will make use of this model to create observable predictions from physical models of multi-phase gas in a wide range of environments.

\section*{Acknowledgements}

RW thanks Erica Nelson for the conversation that sparked the idea for the presented method. We thank the members of the SMAUG collaboration for the many discussions on the physics of multi-phase gases and the anonymous referee for the thoughtful comments and suggestions that greatly improved the quality of this manuscript.

This work was supported by the Natural Sciences and Engineering Research Council of Canada (NSERC), funding reference CITA 490888-16. Some of the computations in this paper were run on the FASRC Cannon cluster supported by the FAS Division of Science Research Computing Group at Harvard University.

\section*{Data Availability}

The data underlying this article will be shared on reasonable request to the corresponding author. The implementation of the numerical method presented in this article is part of a larger project that is planned to be released as open source software upon completion.



\bibliographystyle{mnras}

\begin{thebibliography}{99}
\makeatletter
\relax
\def\mn@urlcharsother{\let\do\@makeother \do\$\do\&\do\#\do\^\do\_\do\%\do\~}
\def\mn@doi{\begingroup\mn@urlcharsother \@ifnextchar [ {\mn@doi@}
  {\mn@doi@[]}}
\def\mn@doi@[#1]#2{\def\@tempa{#1}\ifx\@tempa\@empty \href
  {http://dx.doi.org/#2} {doi:#2}\else \href {http://dx.doi.org/#2} {#1}\fi
  \endgroup}
\def\mn@eprint#1#2{\mn@eprint@#1:#2::\@nil}
\def\mn@eprint@arXiv#1{\href {http://arxiv.org/abs/#1} {{\tt arXiv:#1}}}
\def\mn@eprint@dblp#1{\href {http://dblp.uni-trier.de/rec/bibtex/#1.xml}
  {dblp:#1}}
\def\mn@eprint@#1:#2:#3:#4\@nil{\def\@tempa {#1}\def\@tempb {#2}\def\@tempc
  {#3}\ifx \@tempc \@empty \let \@tempc \@tempb \let \@tempb \@tempa \fi \ifx
  \@tempb \@empty \def\@tempb {arXiv}\fi \@ifundefined
  {mn@eprint@\@tempb}{\@tempb:\@tempc}{\expandafter \expandafter \csname
  mn@eprint@\@tempb\endcsname \expandafter{\@tempc}}}

\bibitem[\protect\citeauthoryear{{Agertz}, {Kravtsov}, {Leitner}  \&
  {Gnedin}}{{Agertz} et~al.}{2013}]{agertz13}
{Agertz} O.,  {Kravtsov} A.~V.,  {Leitner} S.~N.,   {Gnedin} N.~Y.,  2013,
  \mn@doi [\apj] {10.1088/0004-637X/770/1/25}, \href
  {https://ui.adsabs.harvard.edu/abs/2013ApJ...770...25A} {770, 25}

\bibitem[\protect\citeauthoryear{{Applebaum}, {Brooks}, {Christensen},
  {Munshi}, {Quinn}, {Shen}  \& {Tremmel}}{{Applebaum}
  et~al.}{2021}]{applebaum21}
{Applebaum} E.,  {Brooks} A.~M.,  {Christensen} C.~R.,  {Munshi} F.,  {Quinn}
  T.~R.,  {Shen} S.,   {Tremmel} M.,  2021, \mn@doi [\apj]
  {10.3847/1538-4357/abcafa}, \href
  {https://ui.adsabs.harvard.edu/abs/2021ApJ...906...96A} {906, 96}

\bibitem[\protect\citeauthoryear{Baer \& Nunziato}{Baer \&
  Nunziato}{1986}]{baer1986}
Baer M.,  Nunziato J.,  1986, \mn@doi [International Journal of Multiphase
  Flow] {https://doi.org/10.1016/0301-9322(86)90033-9}, 12, 861

\bibitem[\protect\citeauthoryear{{Begelman} \& {Fabian}}{{Begelman} \&
  {Fabian}}{1990}]{begelman90b}
{Begelman} M.~C.,  {Fabian} A.~C.,  1990, \mnras, \href
  {https://ui.adsabs.harvard.edu/abs/1990MNRAS.244P..26B} {244, 26P}

\bibitem[\protect\citeauthoryear{{Begelman} \& {McKee}}{{Begelman} \&
  {McKee}}{1990}]{begelman90}
{Begelman} M.~C.,  {McKee} C.~F.,  1990, \mn@doi [\apj] {10.1086/168994}, \href
  {https://ui.adsabs.harvard.edu/abs/1990ApJ...358..375B} {358, 375}

\bibitem[\protect\citeauthoryear{{Ben{\'\i}tez-Llambay}, {Krapp}  \&
  {Pessah}}{{Ben{\'\i}tez-Llambay} et~al.}{2019}]{benitez-llambay19}
{Ben{\'\i}tez-Llambay} P.,  {Krapp} L.,   {Pessah} M.~E.,  2019, \mn@doi
  [\apjs] {10.3847/1538-4365/ab0a0e}, \href
  {https://ui.adsabs.harvard.edu/abs/2019ApJS..241...25B} {241, 25}

\bibitem[\protect\citeauthoryear{{Borrow}, {Schaller}  \& {Bower}}{{Borrow}
  et~al.}{2021}]{borrow21}
{Borrow} J.,  {Schaller} M.,   {Bower} R.~G.,  2021, \mn@doi [\mnras]
  {10.1093/mnras/stab1423}, \href
  {https://ui.adsabs.harvard.edu/abs/2021MNRAS.505.2316B} {505, 2316}

\bibitem[\protect\citeauthoryear{{Buck}, {Pfrommer}, {Girichidis}  \&
  {Corobean}}{{Buck} et~al.}{2022}]{buck22}
{Buck} T.,  {Pfrommer} C.,  {Girichidis} P.,   {Corobean} B.,  2022, \mn@doi
  [\mnras] {10.1093/mnras/stac952}, \href
  {https://ui.adsabs.harvard.edu/abs/2022MNRAS.513.1414B} {513, 1414}

\bibitem[\protect\citeauthoryear{{Chang} \& {Liou}}{{Chang} \&
  {Liou}}{2007}]{chang07}
{Chang} C.-H.,  {Liou} M.-S.,  2007, \mn@doi [Journal of Computational Physics]
  {10.1016/j.jcp.2007.01.007}, \href
  {https://ui.adsabs.harvard.edu/abs/2007JCoPh.225..840C} {225, 840}

\bibitem[\protect\citeauthoryear{{Cielo}, {Bieri}, {Volonteri}, {Wagner}  \&
  {Dubois}}{{Cielo} et~al.}{2018}]{cielo18}
{Cielo} S.,  {Bieri} R.,  {Volonteri} M.,  {Wagner} A.~Y.,   {Dubois} Y.,
  2018, \mn@doi [\mnras] {10.1093/mnras/sty708}, \href
  {https://ui.adsabs.harvard.edu/abs/2018MNRAS.477.1336C} {477, 1336}

\bibitem[\protect\citeauthoryear{{Dav{\'e}}, {Thompson}  \&
  {Hopkins}}{{Dav{\'e}} et~al.}{2016}]{dave16}
{Dav{\'e}} R.,  {Thompson} R.,   {Hopkins} P.~F.,  2016, \mn@doi [\mnras]
  {10.1093/mnras/stw1862}, \href
  {https://ui.adsabs.harvard.edu/abs/2016MNRAS.462.3265D} {462, 3265}

\bibitem[\protect\citeauthoryear{{Dubois} \& {Teyssier}}{{Dubois} \&
  {Teyssier}}{2008}]{dubois08}
{Dubois} Y.,  {Teyssier} R.,  2008, \mn@doi [\aap]
  {10.1051/0004-6361:20078326}, \href
  {https://ui.adsabs.harvard.edu/abs/2008A&A...477...79D} {477, 79}

\bibitem[\protect\citeauthoryear{{Dubois} et~al.,}{{Dubois}
  et~al.}{2021}]{dubois21}
{Dubois} Y.,  et~al., 2021, \mn@doi [\aap] {10.1051/0004-6361/202039429}, \href
  {https://ui.adsabs.harvard.edu/abs/2021A&A...651A.109D} {651, A109}

\bibitem[\protect\citeauthoryear{{Field}}{{Field}}{1965}]{field65}
{Field} G.~B.,  1965, \mn@doi [\apj] {10.1086/148317}, \href
  {https://ui.adsabs.harvard.edu/abs/1965ApJ...142..531F} {142, 531}

\bibitem[\protect\citeauthoryear{{Fielding}, {Ostriker}, {Bryan}  \&
  {Jermyn}}{{Fielding} et~al.}{2020}]{fielding20}
{Fielding} D.~B.,  {Ostriker} E.~C.,  {Bryan} G.~L.,   {Jermyn} A.~S.,  2020,
  \mn@doi [\apjl] {10.3847/2041-8213/ab8d2c}, \href
  {https://ui.adsabs.harvard.edu/abs/2020ApJ...894L..24F} {894, L24}

\bibitem[\protect\citeauthoryear{{Gatto} et~al.,}{{Gatto}
  et~al.}{2015}]{gatto15}
{Gatto} A.,  et~al., 2015, \mn@doi [\mnras] {10.1093/mnras/stv324}, \href
  {https://ui.adsabs.harvard.edu/abs/2015MNRAS.449.1057G} {449, 1057}

\bibitem[\protect\citeauthoryear{{Girichidis}, {Naab}, {Hanasz}  \&
  {Walch}}{{Girichidis} et~al.}{2018}]{girichidis18}
{Girichidis} P.,  {Naab} T.,  {Hanasz} M.,   {Walch} S.,  2018, \mn@doi
  [\mnras] {10.1093/mnras/sty1653}, \href
  {https://ui.adsabs.harvard.edu/abs/2018MNRAS.479.3042G} {479, 3042}

\bibitem[\protect\citeauthoryear{{Grand} et~al.,}{{Grand}
  et~al.}{2021}]{grand21}
{Grand} R. J.~J.,  et~al., 2021, \mn@doi [\mnras] {10.1093/mnras/stab2492},
  \href {https://ui.adsabs.harvard.edu/abs/2021MNRAS.507.4953G} {507, 4953}

\bibitem[\protect\citeauthoryear{{Gronke}, {Oh}, {Ji}  \& {Norman}}{{Gronke}
  et~al.}{2022}]{gronke21}
{Gronke} M.,  {Oh} S.~P.,  {Ji} S.,   {Norman} C.,  2022, \mn@doi [\mnras]
  {10.1093/mnras/stab3351}, \href
  {https://ui.adsabs.harvard.edu/abs/2022MNRAS.511..859G} {511, 859}

\bibitem[\protect\citeauthoryear{{Hahn} \& {Abel}}{{Hahn} \&
  {Abel}}{2011}]{hahn11}
{Hahn} O.,  {Abel} T.,  2011, \mn@doi [\mnras]
  {10.1111/j.1365-2966.2011.18820.x}, \href
  {https://ui.adsabs.harvard.edu/abs/2011MNRAS.415.2101H} {415, 2101}

\bibitem[\protect\citeauthoryear{{Hopkins}, {Quataert}  \& {Murray}}{{Hopkins}
  et~al.}{2012a}]{hopkins12b}
{Hopkins} P.~F.,  {Quataert} E.,   {Murray} N.,  2012a, \mn@doi [\mnras]
  {10.1111/j.1365-2966.2012.20578.x}, \href
  {https://ui.adsabs.harvard.edu/abs/2012MNRAS.421.3488H} {421, 3488}

\bibitem[\protect\citeauthoryear{{Hopkins}, {Quataert}  \& {Murray}}{{Hopkins}
  et~al.}{2012b}]{hopkins12}
{Hopkins} P.~F.,  {Quataert} E.,   {Murray} N.,  2012b, \mn@doi [\mnras]
  {10.1111/j.1365-2966.2012.20593.x}, \href
  {https://ui.adsabs.harvard.edu/abs/2012MNRAS.421.3522H} {421, 3522}

\bibitem[\protect\citeauthoryear{{Hopkins} et~al.,}{{Hopkins}
  et~al.}{2018}]{hopkins18}
{Hopkins} P.~F.,  et~al., 2018, \mn@doi [\mnras] {10.1093/mnras/sty674}, \href
  {https://ui.adsabs.harvard.edu/abs/2018MNRAS.477.1578H} {477, 1578}

\bibitem[\protect\citeauthoryear{{Huang}, {Katz}, {Scannapieco}, {Cottle},
  {Dav{\'e}}, {Weinberg}, {Peeples}  \& {Br{\"u}ggen}}{{Huang}
  et~al.}{2020}]{huang20}
{Huang} S.,  {Katz} N.,  {Scannapieco} E.,  {Cottle} J.,  {Dav{\'e}} R.,
  {Weinberg} D.~H.,  {Peeples} M.~S.,   {Br{\"u}ggen} M.,  2020, \mn@doi
  [\mnras] {10.1093/mnras/staa1978}, \href
  {https://ui.adsabs.harvard.edu/abs/2020MNRAS.497.2586H} {497, 2586}

\bibitem[\protect\citeauthoryear{{Huang}, {Katz}, {Cottle}, {Scannapieco},
  {Dav{\'e}}  \& {Weinberg}}{{Huang} et~al.}{2022}]{huang22}
{Huang} S.,  {Katz} N.,  {Cottle} J.,  {Scannapieco} E.,  {Dav{\'e}} R.,
  {Weinberg} D.~H.,  2022, \mn@doi [\mnras] {10.1093/mnras/stab3363}, \href
  {https://ui.adsabs.harvard.edu/abs/2022MNRAS.509.6091H} {509, 6091}

\bibitem[\protect\citeauthoryear{{Hummels} et~al.,}{{Hummels}
  et~al.}{2019}]{hummels19}
{Hummels} C.~B.,  et~al., 2019, \mn@doi [\apj] {10.3847/1538-4357/ab378f},
  \href {https://ui.adsabs.harvard.edu/abs/2019ApJ...882..156H} {882, 156}

\bibitem[\protect\citeauthoryear{{Jeans}}{{Jeans}}{1902}]{jeans1902}
{Jeans} J.~H.,  1902, \mn@doi [Philosophical Transactions of the Royal Society
  of London Series A] {10.1098/rsta.1902.0012}, \href
  {https://ui.adsabs.harvard.edu/abs/1902RSPTA.199....1J} {199, 1}

\bibitem[\protect\citeauthoryear{{Kim} \& {Ostriker}}{{Kim} \&
  {Ostriker}}{2017}]{kim17b}
{Kim} C.-G.,  {Ostriker} E.~C.,  2017, \mn@doi [\apj]
  {10.3847/1538-4357/aa8599}, \href
  {https://ui.adsabs.harvard.edu/abs/2017ApJ...846..133K} {846, 133}

\bibitem[\protect\citeauthoryear{{Kim}, {Ostriker}  \& {Raileanu}}{{Kim}
  et~al.}{2017}]{kim17}
{Kim} C.-G.,  {Ostriker} E.~C.,   {Raileanu} R.,  2017, \mn@doi [\apj]
  {10.3847/1538-4357/834/1/25}, \href
  {https://ui.adsabs.harvard.edu/abs/2017ApJ...834...25K} {834, 25}

\bibitem[\protect\citeauthoryear{{Marinacci}, {Sales}, {Vogelsberger}, {Torrey}
   \& {Springel}}{{Marinacci} et~al.}{2019}]{marinacci19}
{Marinacci} F.,  {Sales} L.~V.,  {Vogelsberger} M.,  {Torrey} P.,   {Springel}
  V.,  2019, \mn@doi [\mnras] {10.1093/mnras/stz2391}, \href
  {https://ui.adsabs.harvard.edu/abs/2019MNRAS.489.4233M} {489, 4233}

\bibitem[\protect\citeauthoryear{{McCourt}, {Oh}, {O'Leary}  \&
  {Madigan}}{{McCourt} et~al.}{2018}]{mccourt18}
{McCourt} M.,  {Oh} S.~P.,  {O'Leary} R.,   {Madigan} A.-M.,  2018, \mn@doi
  [\mnras] {10.1093/mnras/stx2687}, \href
  {https://ui.adsabs.harvard.edu/abs/2018MNRAS.473.5407M} {473, 5407}

\bibitem[\protect\citeauthoryear{{Mukherjee}, {Bicknell}, {Wagner},
  {Sutherland}  \& {Silk}}{{Mukherjee} et~al.}{2018}]{mukherjee18}
{Mukherjee} D.,  {Bicknell} G.~V.,  {Wagner} A.~Y.,  {Sutherland} R.~S.,
  {Silk} J.,  2018, \mn@doi [\mnras] {10.1093/mnras/sty1776}, \href
  {https://ui.adsabs.harvard.edu/abs/2018MNRAS.479.5544M} {479, 5544}

\bibitem[\protect\citeauthoryear{{Nelson}, {Genel}, {Pillepich},
  {Vogelsberger}, {Springel}  \& {Hernquist}}{{Nelson} et~al.}{2016}]{nelson16}
{Nelson} D.,  {Genel} S.,  {Pillepich} A.,  {Vogelsberger} M.,  {Springel} V.,
   {Hernquist} L.,  2016, \mn@doi [\mnras] {10.1093/mnras/stw1191}, \href
  {https://ui.adsabs.harvard.edu/abs/2016MNRAS.460.2881N} {460, 2881}

\bibitem[\protect\citeauthoryear{{Nelson} et~al.,}{{Nelson}
  et~al.}{2020}]{nelson20}
{Nelson} D.,  et~al., 2020, \mn@doi [\mnras] {10.1093/mnras/staa2419}, \href
  {https://ui.adsabs.harvard.edu/abs/2020MNRAS.498.2391N} {498, 2391}

\bibitem[\protect\citeauthoryear{{Pakmor}, {Springel}, {Bauer}, {Mocz},
  {Munoz}, {Ohlmann}, {Schaal}  \& {Zhu}}{{Pakmor} et~al.}{2016}]{pakmor16}
{Pakmor} R.,  {Springel} V.,  {Bauer} A.,  {Mocz} P.,  {Munoz} D.~J.,
  {Ohlmann} S.~T.,  {Schaal} K.,   {Zhu} C.,  2016, \mn@doi [\mnras]
  {10.1093/mnras/stv2380}, \href
  {https://ui.adsabs.harvard.edu/abs/2016MNRAS.455.1134P} {455, 1134}

\bibitem[\protect\citeauthoryear{{Peeples} et~al.,}{{Peeples}
  et~al.}{2019}]{peeples19}
{Peeples} M.~S.,  et~al., 2019, \mn@doi [\apj] {10.3847/1538-4357/ab0654},
  \href {https://ui.adsabs.harvard.edu/abs/2019ApJ...873..129P} {873, 129}

\bibitem[\protect\citeauthoryear{{Prosperetti} \& {Tryggvason}}{{Prosperetti}
  \& {Tryggvason}}{2007}]{prosperettiBook07}
{Prosperetti} A.,  {Tryggvason} G.,  2007, {Computational Methods for
  Multiphase Flow}.
Cambridge University Press

\bibitem[\protect\citeauthoryear{{Saurel} \& {Abgrall}}{{Saurel} \&
  {Abgrall}}{1999}]{saurel99}
{Saurel} R.,  {Abgrall} R.,  1999, \mn@doi [Journal of Computational Physics]
  {10.1006/jcph.1999.6187}, \href
  {https://ui.adsabs.harvard.edu/abs/1999JCoPh.150..425S} {150, 425}

\bibitem[\protect\citeauthoryear{{Schaye} \& {Dalla Vecchia}}{{Schaye} \&
  {Dalla Vecchia}}{2008}]{schaye08}
{Schaye} J.,  {Dalla Vecchia} C.,  2008, \mn@doi [\mnras]
  {10.1111/j.1365-2966.2007.12639.x}, \href
  {https://ui.adsabs.harvard.edu/abs/2008MNRAS.383.1210S} {383, 1210}

\bibitem[\protect\citeauthoryear{{Schneider} et~al.,}{{Schneider}
  et~al.}{2016}]{schneider16}
{Schneider} A.,  et~al., 2016, \mn@doi [\jcap] {10.1088/1475-7516/2016/04/047},
  \href {https://ui.adsabs.harvard.edu/abs/2016JCAP...04..047S} {2016, 047}

\bibitem[\protect\citeauthoryear{{Semenov}, {Kravtsov}  \& {Gnedin}}{{Semenov}
  et~al.}{2017}]{semenov17}
{Semenov} V.~A.,  {Kravtsov} A.~V.,   {Gnedin} N.~Y.,  2017, \mn@doi [\apj]
  {10.3847/1538-4357/aa8096}, \href
  {https://ui.adsabs.harvard.edu/abs/2017ApJ...845..133S} {845, 133}

\bibitem[\protect\citeauthoryear{{Simpson}, {Pakmor}, {Marinacci}, {Pfrommer},
  {Springel}, {Glover}, {Clark}  \& {Smith}}{{Simpson}
  et~al.}{2016}]{simpson16}
{Simpson} C.~M.,  {Pakmor} R.,  {Marinacci} F.,  {Pfrommer} C.,  {Springel} V.,
   {Glover} S. C.~O.,  {Clark} P.~C.,   {Smith} R.~J.,  2016, \mn@doi [\apjl]
  {10.3847/2041-8205/827/2/L29}, \href
  {https://ui.adsabs.harvard.edu/abs/2016ApJ...827L..29S} {827, L29}

\bibitem[\protect\citeauthoryear{{S{\k{a}}dowski}, {Wielgus}, {Narayan},
  {Abarca}, {McKinney}  \& {Chael}}{{S{\k{a}}dowski} et~al.}{2017}]{sadowski17}
{S{\k{a}}dowski} A.,  {Wielgus} M.,  {Narayan} R.,  {Abarca} D.,  {McKinney}
  J.~C.,   {Chael} A.,  2017, \mn@doi [\mnras] {10.1093/mnras/stw3116}, \href
  {https://ui.adsabs.harvard.edu/abs/2017MNRAS.466..705S} {466, 705}

\bibitem[\protect\citeauthoryear{{Smith}, {Sijacki}  \& {Shen}}{{Smith}
  et~al.}{2018}]{smith18}
{Smith} M.~C.,  {Sijacki} D.,   {Shen} S.,  2018, \mn@doi [\mnras]
  {10.1093/mnras/sty994}, \href
  {https://ui.adsabs.harvard.edu/abs/2018MNRAS.478..302S} {478, 302}

\bibitem[\protect\citeauthoryear{{Somerville} \& {Dav{\'e}}}{{Somerville} \&
  {Dav{\'e}}}{2015}]{somerville15}
{Somerville} R.~S.,  {Dav{\'e}} R.,  2015, \mn@doi [\araa]
  {10.1146/annurev-astro-082812-140951}, \href
  {https://ui.adsabs.harvard.edu/abs/2015ARA&A..53...51S} {53, 51}

\bibitem[\protect\citeauthoryear{{Springel}}{{Springel}}{2010}]{springel10}
{Springel} V.,  2010, \mn@doi [\mnras] {10.1111/j.1365-2966.2009.15715.x},
  \href {https://ui.adsabs.harvard.edu/abs/2010MNRAS.401..791S} {401, 791}

\bibitem[\protect\citeauthoryear{{Springel} \& {Hernquist}}{{Springel} \&
  {Hernquist}}{2003}]{springel03}
{Springel} V.,  {Hernquist} L.,  2003, \mn@doi [\mnras]
  {10.1046/j.1365-8711.2003.06206.x}, \href
  {https://ui.adsabs.harvard.edu/abs/2003MNRAS.339..289S} {339, 289}

\bibitem[\protect\citeauthoryear{{Veilleux}, {Maiolino}, {Bolatto}  \&
  {Aalto}}{{Veilleux} et~al.}{2020}]{veilleux20}
{Veilleux} S.,  {Maiolino} R.,  {Bolatto} A.~D.,   {Aalto} S.,  2020, \mn@doi
  [\aapr] {10.1007/s00159-019-0121-9}, \href
  {https://ui.adsabs.harvard.edu/abs/2020A&ARv..28....2V} {28, 2}

\bibitem[\protect\citeauthoryear{{Vogelsberger} et~al.,}{{Vogelsberger}
  et~al.}{2014}]{vogelsberger14}
{Vogelsberger} M.,  et~al., 2014, \mn@doi [\nat] {10.1038/nature13316}, \href
  {https://ui.adsabs.harvard.edu/abs/2014Natur.509..177V} {509, 177}

\bibitem[\protect\citeauthoryear{{\VAN{Voort}{van de}{van de} Voort},
  {Springel}, {Mandelker}, {van den Bosch}  \& {Pakmor}}{{\VAN{Voort}{van
  de}{van de} Voort} et~al.}{2019}]{vandevoort19}
{\VAN{Voort}{van de}{van de} Voort} F.,  {Springel} V.,  {Mandelker} N.,  {van
  den Bosch} F.~C.,   {Pakmor} R.,  2019, \mn@doi [\mnras]
  {10.1093/mnrasl/sly190}, \href
  {https://ui.adsabs.harvard.edu/abs/2019MNRAS.482L..85V} {482, L85}

\bibitem[\protect\citeauthoryear{{Wagner}, {Bicknell}  \& {Umemura}}{{Wagner}
  et~al.}{2012}]{wagner12}
{Wagner} A.~Y.,  {Bicknell} G.~V.,   {Umemura} M.,  2012, \mn@doi [\apj]
  {10.1088/0004-637X/757/2/136}, \href
  {https://ui.adsabs.harvard.edu/abs/2012ApJ...757..136W} {757, 136}

\bibitem[\protect\citeauthoryear{{Walch} et~al.,}{{Walch}
  et~al.}{2015}]{walch15}
{Walch} S.,  et~al., 2015, \mn@doi [\mnras] {10.1093/mnras/stv1975}, \href
  {https://ui.adsabs.harvard.edu/abs/2015MNRAS.454..238W} {454, 238}

\bibitem[\protect\citeauthoryear{{Weinberger}, {Springel}  \&
  {Pakmor}}{{Weinberger} et~al.}{2020}]{weinberger20}
{Weinberger} R.,  {Springel} V.,   {Pakmor} R.,  2020, \mn@doi [\apjs]
  {10.3847/1538-4365/ab908c}, \href
  {https://ui.adsabs.harvard.edu/abs/2020ApJS..248...32W} {248, 32}

\bibitem[\protect\citeauthoryear{{Yee}, {Vinokur}  \& {Djomehri}}{{Yee}
  et~al.}{2000}]{yee00}
{Yee} H.~C.,  {Vinokur} M.,   {Djomehri} M.~J.,  2000, \mn@doi [Journal of
  Computational Physics] {10.1006/jcph.2000.6517}, \href
  {https://ui.adsabs.harvard.edu/abs/2000JCoPh.162...33Y} {162, 33}

\makeatother
\end{thebibliography}




\appendix

\section{Resolved fluid Instabilities}
\subsection{Kelvin-Helmholtz instability}
\begin{figure}
    \centering
    \includegraphics{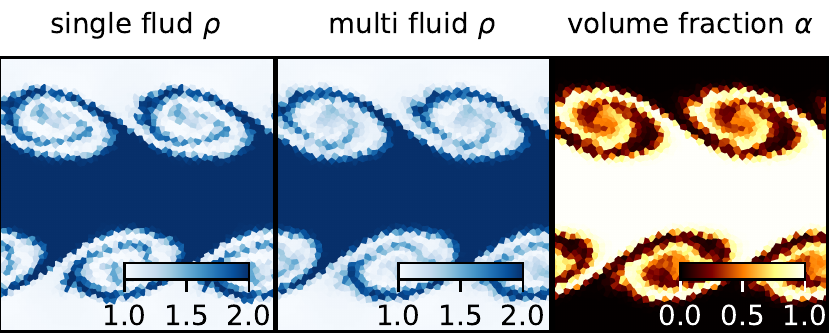}
    \caption{Left panel: density of single-fluid Kelvin-Helmholtz instability. Center: volume averaged density of the equivalent 2-fluid setup. Right panel: volume fraction.}
    \label{fig:khi}
\end{figure}

In the case of spatially resolved multi-phase structures a key criterion for the quality of the modeling is the ability to handle fluid instabilities identically to the single-fluid case. We test this via Kelvin-Helmholtz (KH) and Rayleigh-Taylor (RT) instabilities in which a single spatial mode is seeded. We adopt the examples from \cite{springel10}, and generalize the setup to the 2-fluid case by choosing a volume fraction of $10^{-2}$ and $(1-10^{-2})$ depending on the initial density. This implies that the phase boundary in the 2-fluid simulation is mainly dependent on the cross-fluid terms is subject to advection errors in the volume-fraction. Since both phases are identical to the single fluid one, we expect the same result, except for the aforementioned numerical aspects.

The KH instability is set up in a periodic 2d box of side length $1$ discretized by $50$ cells per dimension. The coordinates range from $0<x<1$ and $0<y<1$. The background has  the following density $\rho$, velocity $\vec{v}$ and pressure $p$:
\begin{align}
    \rho &= 1 \nonumber\\
    \vec{v} &= \left(-0.5,\,w\,\sin\left(4\,\pi x\right)\right)^{T} \nonumber\\
    w &= 0.1 \left(\exp\left(-\frac{(y-0.25)^2}{0.0025} \right) +  \exp\left(-\frac{(y-0.75)^2}{0.0025} \right) \right) \\
    p &= 2.5. \nonumber
\end{align}
In the central region, $0.25<y<0.75$, we set up a band of denser gas moving in the opposite direction,
\begin{align}
    \rho &= 2.0 \nonumber \\
    \vec{v} &= \left(0.5,\,w\,\sin\left(4\,\pi x\right)\right)^{T}\\
    p &= 2.5. \nonumber
\end{align}
The fluid has an adiabatic index $\gamma=1.4$ throughout, and a CFL factor of $0.3$ is used.

Fig.~\ref{fig:khi} shows the result at time $t=2.0$. The left panel is the density in the single fluid simulation, the center panel the volume-averaged density $\rho = \alpha\rho_1 + (1-\alpha)\rho_2$ of the 2-fluid simulation, and the volume fraction $\alpha$ in the right panel. Most notably, the density slices are similar in overall shape, showing that the cross-term accurately captures resolved fluid-instabilities. Looking closely on a cell-level basis in individual KH eddies, the single fluid case exhibits sharper contrasts between cells while the multifluid case shows mixing on slightly larger scales and consequently a more smooth appearance. This is likely due to mixing of the volume-fractions and expected from the relatively large advection errors. Overall, however, we find an unaltered performance except for length scales comparable to individual cell sizes.

\subsection{Rayleigh-Taylor instability}

\begin{figure}
    \centering
    \includegraphics{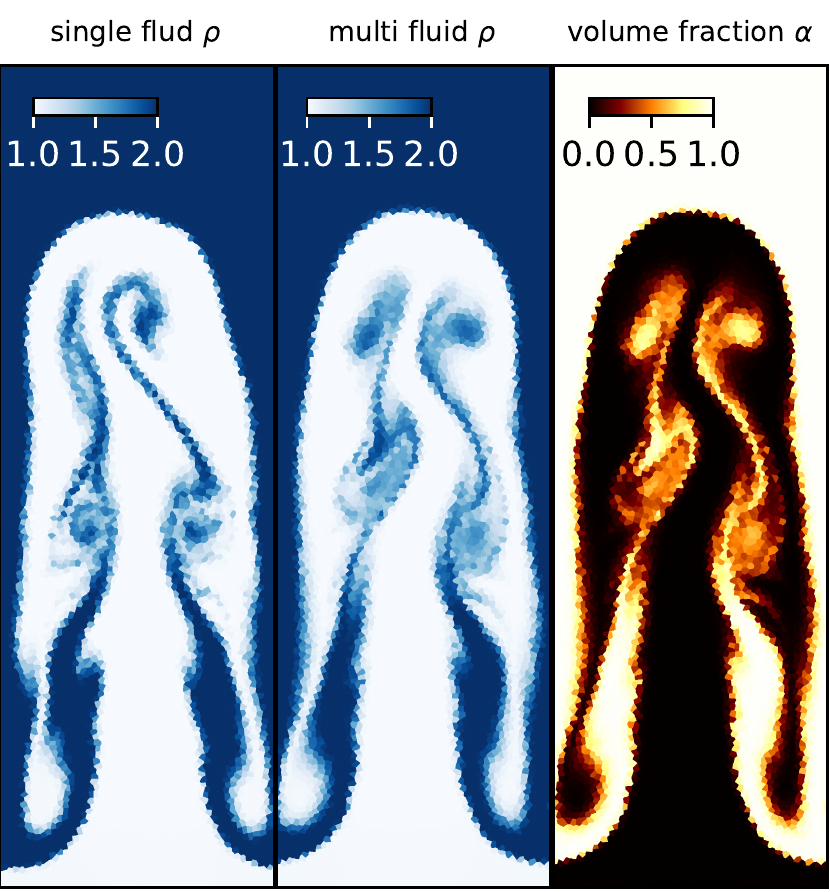}
    \caption{Left panel: density of single-fluid Rayleigh-Taylor instability. Center: volume averaged density of the equivalent 2-fluid setup. Right panel: volume fraction.}
    \label{fig:rti}
\end{figure}

In order to test the coupling to a gravitational acceleration $g=-0.1$, we carry out a simulation of a RT instability, again with the setup of \citet{springel10} and its 2-fluid generalization. Concretely, we set up a 2d box with $0<x<0.5$ and $0<y<1.5$ discretized with $48\times 144$ cells. The initial conditions are
\begin{align}
    \rho & =
\begin{cases}
    1.0 \quad\text{if}y<0.75\\
    2.0 \quad\text{if}y\geq0.75\\
\end{cases} \nonumber\\
    \vec{v} &= \left(0.0,\, 0.0025\left[1-\cos\left( 4 \pi x\right)\right] \left[1-\cos\left( 4 \pi y / 3\right)\right]\right)^{T}\\
    p &= 2.5 - g\,(y-0.75) \,\rho . \nonumber
\end{align}
The adiabatic index is $\gamma = 1.4$ and the employed CFL factor $0.3$. As in the KH instability case, the volume fraction follows the initial densities, with $\alpha=10^{-2}$ and $\alpha=(1-10^{-2})$.

Fig.~\ref{fig:rti} shows the result at $t=15$.  On the left is the density for the single fluid simulation, in the center the equivalent volume-averaged density in the 2-fluid simulation, and on the right is the volume fraction $\alpha$. Similarly to the KH case, the overall behaviour is very similar, with slightly more mixing on spatial scales of individual cells. The precise morphology of the phases are somewhat different, however, most likely due minor differences in the mesh movements which ultimately breaks the symmetry in this problem \citep[see][section 8.8, for a detailed discussion on this topic]{springel10}. Most notably, however, the differences due to 2-fluid modeling are smaller than the differences due to details of the discretization.


\bsp	
\label{lastpage}
\end{document}